\RequirePackage{etex}
\documentclass{aa}  

\usepackage{graphicx}
\usepackage{textcomp}
\usepackage{txfonts}
\usepackage{natbib}
\usepackage{xcolor}

\usepackage{gensymb}
\usepackage{graphicx}
\usepackage{txfonts}
\usepackage{lipsum}
\usepackage{subcaption}         
\usepackage{lscape}             
\usepackage{placeins}           
                                
\begin{document}

   \title{Near-infrared periodicity in the SPICY catalog}

   \subtitle{True young stellar objects versus contaminants}

   \titlerunning{NIR periodicity in the SPICY catalog}

%

   \author{C. Ordenes-Huanca\inst{1,2}\thanks{Corresponding author: \email{ccordenes@uc.cl}}
        \and A. Bayo\inst{3}
        \and M. Zoccali\inst{1}
        \and V. Guzmán\inst{1,6}
        \and A. Rojas-Arriagada\inst{4,5}
        }

   \institute{Instituto de Astrofísica, Pontificia Universidad Católica de Chile, Casilla 306, Santiago 7820436, Chile
   \and Departamento de Astronomía, Universidad de Concepción, Casilla 160-C, Concepción, Chile 
   \and European Southern Observatory, Karl-Schwarzschild-Strasse 2, 85748 Garching bei München, Germany
   \and Departamento de Física, Universidad de Santiago de Chile, Av. Víctor Jara 3659, 9170124, Santiago, Chile
   \and Center for Interdisciplinary Research in Astrophysics and Space Exploration (CIRAS), Universidad de Santiago de Chile, Santiago, Chile
   \and Millennium Nucleus on Young Exoplanets and their Moons (YEMS), Chile}

   \date{Received 2 April 2026 / Accepted 29 September 2026}

 
  \abstract
   {Infrared (IR) large-sky surveys allow study of how localized star formation is, enabling searches for young stellar objects (YSOs) not only at the centers of known stellar nurseries but also across the Galactic plane. However, these spatially unbiased searches suffer from contamination, most notably from dusty asymptotic giant branch (AGB) stars characterized by long-period variability.}
   {Using time-series analysis, and taking advantage of the different variability timescales of YSOs against long-period variables (LPVs), we assess the level of contamination of one of the most spatially unbiased YSO catalogs (\textit{Spitzer}/IRAC Candidate YSO, SPICY). Distinguishing between these two groups lets us draw more robust conclusions about the frequency of isolated YSOs in our Galaxy and their implications.} 
   {We cross-match the SPICY and VIRAC2 catalogs (VVV/VVVX Survey), build near-IR light curves for common sources, and compute their periods and amplitudes. Contaminants are identified as sources with periodic flux variations with periods $P\geq 80\,d$ and amplitudes $\Delta K_{\rm s}\geq0.35\,\rm{mag}$.}
   {From $58\,737$ common sources, we identify $742$ SPICY objects as LPV contaminants based on their long-period variability, and identify $307$ high-confidence periodic YSO candidates exhibiting short-period flux changes consistent with cool spots. Near- and mid-IR amplitudes are correlated only among LPVs, suggesting a common pulsation origin. In addition, the spatial distribution of the identified populations further supports our classification, with LPVs preferentially associated with the bulge and YSOs at the lowest galactic latitudes.}
  {}


   \maketitle
   \nolinenumbers
\section{Introduction}

Infrared (IR) observations allow us to pierce through extinction, identifying and characterizing populations intrinsically reddened (via envelopes, disks, or dusty atmospheres) or extrinsically reddened. In particular, space missions like \textit{Spitzer} \citep{Spitzer_2004} and WISE \citep{WISE_2010} have mapped the most obscured parts of our Galaxy, where most star-forming regions are located. Despite major advances, fundamental aspects of star formation in the Milky Way remain debated. The classical picture holds that stars form predominantly in compact, embedded clusters within molecular clouds \citep{Lada_2003}. As the most massive members of these clusters evolve rapidly, their winds and supernova feedback drive violent gas expulsion, leaving most clusters gravitationally unbound within a few million years of their birth. Those that survive this phase gradually dissolve into the Galactic disc over timescales of $\lesssim 1\,\rm Gyr$ through interactions with giant molecular clouds, spiral arms, and the Galactic tidal field \citep{Lamers_2006}. An alternative picture has emerged in which most stars are born in low-density, gravitationally unbound associations that mirror the fractal structure of their natal gas \citep{Ward_2020}. Whether such hierarchical structures represent a genuinely distinct formation pathway or are simply the dissolving remnants of clustered star formation remains an open question \citep{Quintana_2025}.\\

Understanding the spatial distribution of young stars in the Galactic plane provides discriminatory diagnostics, but requires careful sample selection. \citet{Kuhn_2021} did not limit the search to clustered young stars in star-forming complexes; instead, they sought these objects across the full extent of the inner $613\,\mathrm{deg^2}$ of the Galactic mid-plane. Although the method used by the authors of the SPICY catalog to identify young stars is well suited to assessing whether stars form in clustered or isolated environments, it also introduces contaminants into the sample. More evolved sources, particularly thermally pulsating AGB stars, can mimic YSO colors due to their temperatures and IR excess. These stars can undergo luminosity-correlated pulsations \citep{Catelan_2015}, and their mass-loss processes generate circumstellar envelopes with colors similar to those observed in YSOs. \textit{Spitzer} colors have been used to inform classical cuts in color--color diagrams \citep{Allen_2004, Rebull_2010}, but significant overlap exists between AGB stars and YSOs in that parameter space. To overcome this limitation, researchers have developed machine learning classifiers that exploit the full multi-wavelength parameter space simultaneously \citep{Lakshmipathaiah_2023, Jheonn_2026}, but these photometric methods still fail in regions of overlapping parameter space.\\

X-ray observations can help separate these two groups. It has been known that stellar X-ray emission is inversely correlated with age \citep{FEIGELSON_1982, Preibisch_2005, Ramstedt_2011}. Due to their evolutionary stage and the physical process experienced by forming stars, YSOs are known to be strong emitters at this wavelength range. Their X-ray emission primarily comes from accretion shocks and flares driven by intense magnetic activity \citep{Feigelson_1999}. In contrast, long-period variables (LPVs), or dust-enshrouded AGB stars, are not expected to be significant X-ray emitters due to their more advanced evolutionary stage, unless they are members of X-ray binary systems \citep{Ramstedt_2011, Ortiz_2021}. Unfortunately, to date, no large-scale high-spatial resolution X-ray survey covers the plane of our Galaxy.\\

Alternatively, variability analysis provides another avenue to distinguish between YSOs and AGB stars, as their light curves exhibit fundamentally different behaviors. Pulsating AGB stars are often referred to as LPVs with periods of the order of tens to thousands of days and amplitudes typically between $0.3$--$0.8\,\mathrm{mag}$ in $K_\mathrm{s}$ \citep{Whitelock_2000, Matsunaga_2009, Nikzat_2022}. This class includes Mira and semi-regular variables, both located in the AGB evolutionary stage. Mira variables are cool giant stars near the tip of the AGB. Due to their radial pulsations, they show very regular and periodic variation patterns in timescales of $80$--$1\,000\,\mathrm{days}$. Semi-regular variables, on the other hand, are pulsating red giant stars on the AGB. Stars belonging to one of the semi-regular subclasses show periodic flux changes. Still, their light curve shapes are evolving in time with periods ranging from $20$--$1000\,\rm{days}$ and their $K_{\rm{s}}$ amplitudes are expected to be smaller than those of Miras \citep{Catelan_2015, Trabucchi_2021}. In addition, thermally-pulsing AGB stars can also show irregular flux variations, which means that not all stars in this phase exhibit periodic variability \citep{Lu_2025}. For example, for a stellar population of $\sim 12\,$Gyr, solar metallicity, and luminosity of $10^{5}\,L_{\odot}$, it is expected that only a fraction of $0.0625$ stars of their population are in the thermally pulsing phase. LPVs account for an even smaller fraction \citep{Greggio_2011}.\\

In contrast, periodic variability in YSOs arises primarily from magnetic spots coupled with stellar rotation \citep{Bouvier_1995, Henderson_2012}. They typically exhibit periods ranging from several to tens of days, with variability amplitudes of $0.1$--$0.2\,\mathrm{mag}$ in the $K_\mathrm{s}$ band \citep{OrdenesHuanca_2022}. We have observed such periodic variations across multiple wavelengths. Still, because both populations are primarily detected in the IR, mid-IR time-series studies have proven particularly effective for constraining their variability properties \citep{Rebull_2014, Park_2021}. Recent work has demonstrated that time-series data can effectively separate these groups \citep{Jheonn_2026}, establishing light curve analysis as a powerful tool for distinguishing YSOs from AGB contaminants.\\

Optical light curves have been used to identify AGB stars in the thermally pulsing phase, considering specific characteristics such as the amplitude of the flux changes and the measured periods \citep{Lu_2025}. Optical surveys, on the other hand, have been an important tool to identify LPVs, such as the Optical Gravitational Lensing Experiment \citep[OGLE,][]{Soszynski_2009} and the All-Sky Automated Survey for Supernovae \citep[ASAS-SN,][]{Jayasinghe_2019}. More recently, \textit{Gaia} DR3 has enabled the finding of millions of variable stars in optical bands, including LPV candidates (\citeauthor{Sanders_2023} \citeyear{Sanders_2023}; \citeauthor{Lebzelter_2023} \citeyear{Lebzelter_2023}; \citeauthor{MaizApellaniz_2023} \citeyear{MaizApellaniz_2023}). Due to their large amplitude variations, Miras could be identified in several variable star catalogs \citep{Bhardwaj_2019}. These optical surveys are, however, limited by extinction, but their IR counterparts offer similar diagnostics and are not limited by the red nature of AGB stars.\\

Recently, \citet{Jheonn_2026} applied several classifiers (namely, a support vector machine, a random forest, and a multilayer perceptron) to Wide-field Infrared Survey Explorer \citep[hereafter WISE,][]{WISE_2010} $+$ NEOWISE light curves (in $W1$ and $W2$ bands, $3.4\,\mathrm{\mu m}$ and $4.6\,\mathrm{\mu m}$, respectively) to distinguish between YSOs and AGB stars. The WISE+NEOWISE light curves cover a $10$ year baseline ($2013$--$2023$) with sparse observations ($\sim20$ epochs), and a moderate spatial resolution of $\sim 6\,\rm{arcsec}$ at $W1$ wavelengths. These characteristics provide discriminatory power between LPVs and YSOs but are severely limited by angular resolution and cadence.\\

While space-based surveys provide valuable data, ground-based surveys are not to be dismissed. In this context, the Vista Variables of the Vía Láctea (VVV) and its extension (VVVX) surveys \citep{minniti+2010, Saito_2024} are among the most unique surveys of their kind. Using the $K_\mathrm{s}$-band in the near-IR, it monitored the Bulge and part of the disk of the Milky Way for approximately twelve years. Its last data release, the VVV Infrared Astrometric catalogs (VIRAC) Version 2 \citep[VIRAC2,][]{Smith_2025}, takes its first nine years of observations to obtain point-spread-function photometry in $ZYJHK_{\rm{s}}$ bands with an angular resolution of $\sim 1\,\rm{arsec}$, along with $K_{\rm{s}}$ time-series of more than $545$ million stars. In addition, its main source catalog also provides parallaxes and proper motions for a subset of these objects and reaches depths of up to $K_{\rm{s}}\approx 17.5\,\rm{mag}$. Light curves in this catalog, on the other hand, contain on the order of $\sim 100$ epochs of seeing-limited observations (with seeing $< 2\,\rm{arcsec}$). This makes the data sensitive to brightness changes on timescales of years, making it well suited for identifying individual stars and LPVs \citep{Nikzat_2022, Guo_2022}. In the context of YSO studies, these near-IR light curves are especially valuable because they provide periods, amplitudes, and light curve morphologies that can help distinguish between YSO variability and contaminants such as LPVs.\\

Spatially unbiased searches for YSOs across the Galactic plane, such as the SPICY catalog, are powerful tools for constraining the relative importance of these formation pathways. However, the reliability of conclusions drawn from such catalogs depends critically on the purity of the YSO sample. This is particularly relevant because Galactic star formation rate estimates based on YSO counts \citep{Robitaille_2010} already disagree with standard extragalactic indicators \citep{Heiderman_2010, Chomiuk_2011}. Clean, unbiased samples can address this tension. Ensuring the purity of YSO samples is therefore crucial for interpreting star formation rates and distributions. This study therefore aims to identify and remove pulsating AGB contaminants from the SPICY catalog and quantify their spatial distribution across the inner Galactic mid-plane. At the same time, we strengthen the classification of genuine YSO candidates by showing that their periods provide additional evidence of their young nature.

\section{Analysis}

\subsection{Period search}
\label{sec:period_search}

The SPICY catalog ($\sim120\,000$ YSO candidates) was X-matched against the VIRAC2 source catalog ($545$ million objects) with a “best-match” strategy within $1\,\rm{arcsec}$ radius, resulting in a base catalog of $58\,737$ sources. To this base catalog, we applied the following photometric and astrometric criteria: \texttt{ast\_res\_chisq} $<30$, \texttt{chi} $<5$ \citep{Smith_2025}. These cuts yielded a curated catalog of $54\,580$ light curves.\\

We performed the period search on sources with at least $50$ usable $K_{\rm s}$-band epochs and further required the mean photometric uncertainty to be smaller than $30\%$ of the measured raw peak-to-peak amplitude, $\Delta K_{s,\rm raw} = K_{s,\max} - K_{s,\min}$. We used this raw amplitude only as a preliminary period-search quality criterion; we define the robust amplitude used later separately. For these sources, we computed their Lomb--Scargle periodograms \citep{Lomb1976, Scargle1982} considering a period range between $0.4$--$1333\,\mathrm{days}$. The breakup velocity of a typical T Tauri star, about $0.4\,\rm{days}$ \citep{Bertout_1989}, is related to the lower limit in the period range, whereas the upper limit of $1\,333\,\rm{days}$ is related to the baseline of VIRAC2 observations, which are almost one third of the entire follow-up.\\

Periods close to the main sampling aliases at $0.5$, $1$, and $2\,\rm{days}$ were masked using a tolerance of $\pm0.02\,\rm days$. We required each source to have at least two non-alias periodogram peaks after this masking step. To ensure the retained peaks correspond to distinct periodic solutions, we required secondary peaks to be separated from stronger peaks in period space. Specifically, after selecting a peak at period $P_i$, additional peaks within a window of width $\max(0.05~{\rm days}, 0.2P_i)$ were not considered independent. This procedure prevents multiple nearby peaks from the same broad periodogram feature from being counted as separate candidate periods, while still retaining physically distinct long-period solutions. These criteria produced a period search table of $45\,499$ sources.\\

Since we are interested only in true periodic variables, we considered the periodicity $Q$ and asymmetry $M$ parameters defined in \citet{Cody_2014} and adapted for VVVX data in \citet{OrdenesHuanca_2022}. The former quantifies the stability of the detected period. In contrast, the latter quantifies the level of asymmetry in the flux distribution over time, i.e., whether the object is more commonly at a “high level state, a “low level” state, or alternates between higher and lower level states in a balanced manner. Together, these metrics guide classification of light curve variability. The adopted period for the automatic classification was chosen as the candidate period that minimized $Q$, corresponding to the phase-folded solution with the smallest residual scatter after subtracting the smoothed periodic component.\\

For a given trial period $P_{i}$, the periodicity $Q$ was computed as follows. We first removed outliers from the raw $K_{\rm s}$ light curve using a median absolute deviation (MAD) clip at $10\sigma$ around the median magnitude, retaining only sources with at least $50$ epochs. We then phase-folded the light curve and required at least $10$ of $20$ equally spaced phase bins to contain at least one epoch; otherwise, $Q$ was considered undefined for that trial period. We smoothed the phase-folded light curve using a wrapped (circular) boxcar moving average with a window equal to $10\%$ of the number of epochs, and computed the residuals between the smoothed model and the data. We rejected points deviating by more than $3\sigma$ from the mean residual as outliers relative to the phase-folded model, then repeated the smoothing step on the cleaned light curve to obtain the final residuals. The periodicity metric was then defined as
\begin{equation}
    Q = \frac{\operatorname{Var}(\mathrm{resid}) - \sigma_{\mathrm{ph}}^2}
             {\operatorname{Var}(K_s) - \sigma_{\mathrm{ph}}^2},
\end{equation}
where $\operatorname{Var}(\mathrm{resid})$ is the variance of the residuals with respect to the smoothed periodic model, $\operatorname{Var}(K_{\rm s})$ is the total variance of the cleaned light curve, and $\sigma_{\mathrm{ph}}^2$ is the median squared photometric uncertainty of the retained epochs. By construction, $Q \to 0$ for a light curve that describes a perfectly periodic pattern, and $Q \to 1$ is related to a light curve with no periodic signal above the photometric noise. Only physically defined $Q$ values ($0<Q<1$) were retained for subsequent analysis. We applied the same $Q$ definition consistently across the period selection, grid-convergence, and long-period refinement steps described below.\\

The asymmetry parameter, $M$, was computed on the same cleaned light curve used to compute $Q$ (Sect.~\ref{sec:period_search}), restricted to the interval $55\,500 \leq \mathrm{MJD} \leq 57\,000$, which corresponds to the highest-cadence observing window in the VVV/VVVX data. Following \citet{Cody_2014}, $M$ was defined as the difference between the mean magnitude of the epochs lying outside the $10$th--$90$th percentile range and the median magnitude, normalized by the standard deviation of the light curve within this window.\\

For each source, we retained the two strongest independent Lomb--Scargle periods, hereafter $P_1$ and $P_2$, and computed the $Q$ metric for the phase-folded light curve associated with each candidate period. We defined the adopted period as the candidate period that minimized $Q$. This step is necessary because the strongest Lomb--Scargle peak does not always provide the most coherent phase-folded solution. Both $Q(P_1)$ and $Q(P_2)$ were physically defined ($0<Q<1$) for $37\,046$ sources. In $67.9\%$ of these cases, $P_1$ yielded the lowest $Q$ value. In the remaining $32.1\%$ cases, the secondary independent period $P_2$ yielded a lower $Q$ value and was therefore preferred by the automatic period selection step.\\

\begin{figure}
        \centering
        \includegraphics[width=0.8\hsize]{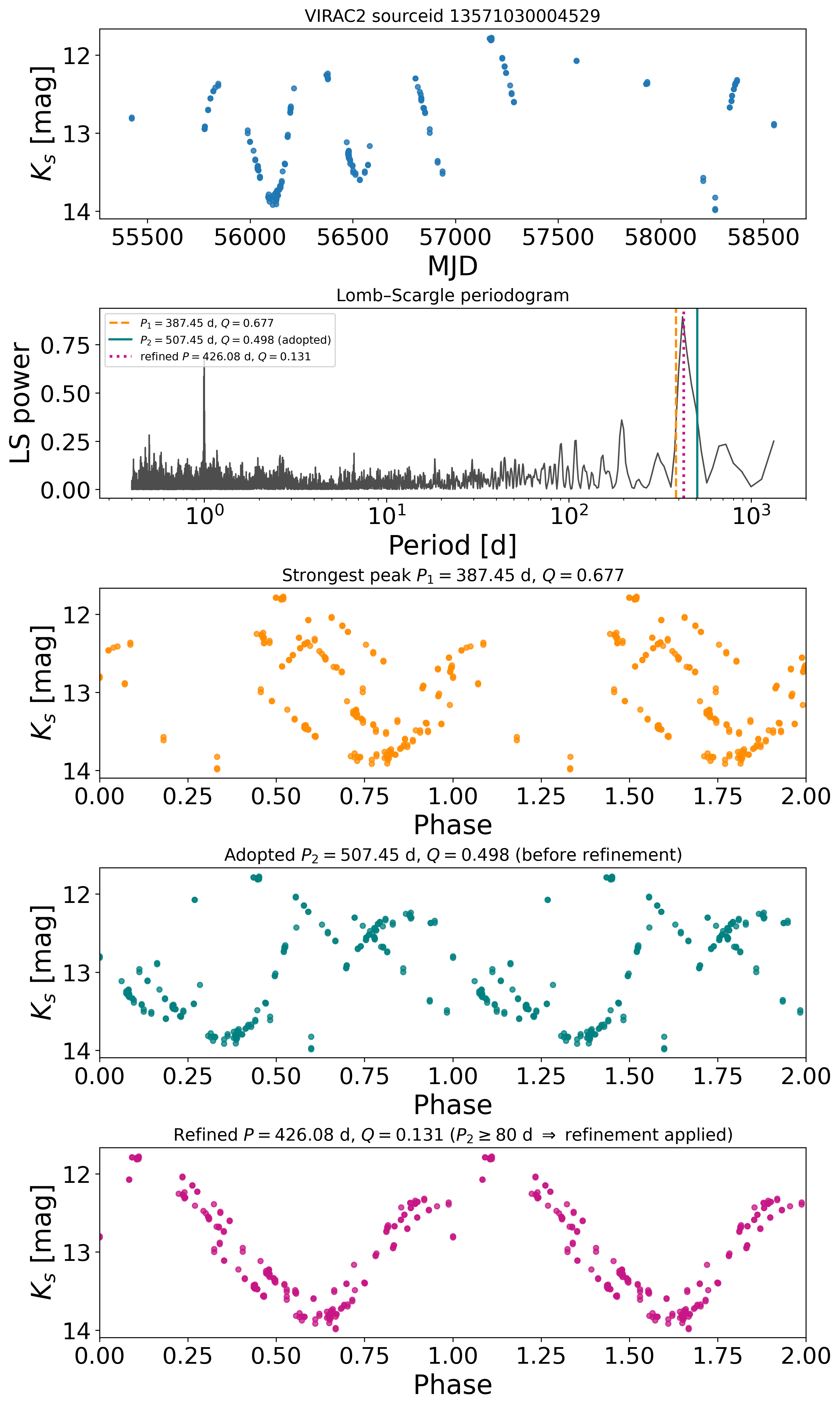}
        \caption{Example illustrating both $P_1$/$P_2$ selection and long-period refinement for VIRAC2 source 13571030004529. \textit{Top:} raw $K_s$ light curve. \textit{Second panel:} Lomb--Scargle periodogram, with the strongest peak ($P_1$, dashed), the adopted second peak ($P_2$, solid), and the refined period (dotted) marked. \textit{Panels 3--5:} phase-folded light curve using $P_1$, the adopted $P_2$ (before refinement), and the refined period, respectively.}
        \label{fig:combined_p1p2_refinement}%
\end{figure}

For stars in which the period that minimized its periodicity was $P\geq 80\,\rm days$, we subsequently refined the period value. For these variables, the periodogram peaks can be broad when expressed in period space, because a uniform frequency spacing implies a period resolution that scales approximately as $\Delta P \propto P^2 \Delta f$ (see the second panel of the example shown in Fig.~\ref{fig:combined_p1p2_refinement}). As a result, small frequency shifts around a long-period peak can correspond to large period shifts while still tracing the same broad periodogram feature. The refinement process involved building a local frequency grid around the selected period and testing the corresponding half- and double-period windows when allowed by the period limits. At each trial period, we phase-folded the light curve and recomputed $Q$. We defined the refined period as the trial period that minimized $Q$ within these local windows. We adopted the refined period only when it lowered $Q$ by at least $0.03$. This refinement was applied to $22\,205$ sources with $P \geq 80$ days; $17\,405$ of them ($78.4\%$) satisfied a physically valid improvement. An example of the resulting refined period is shown in the lower phase-folded panel of Fig.~\ref{fig:combined_p1p2_refinement}, where the adopted period ($P_2$) is further refined from $507.45$ to $426.08\,\rm days$, improving $Q$ from $0.498$ to $0.131$.\\

An example of phase-folded light curves, for the same star, using these different period values is shown in Fig.~\ref{fig:combined_p1p2_refinement}. The upper phase-folded panel shows the light curve folded with $P_1$ (panel with orange dots), the strongest Lomb--Scargle peak, where the pattern is unclear. The middle panel shows the same light curve folded with $P_2$, which is the period that minimizes $Q$ for this source (panel with dark green dots). Here, a periodic modulation becomes clearer, although the phase is not perfectly consistent across cycles. This comparison highlights the need to evaluate the quality of the phase-folded light curve, rather than relying exclusively on a single automatic period estimate.\\

However, since we considered a fixed frequency grid for the Lomb--Scargle analysis, the measured $Q$ value could be affected by the resolution of this grid. Therefore, we made a grid-convergence check by repeating the Lomb--Scargle period search using baseline-dependent uniform frequency grids with $\Delta f = 1/(T \times \texttt{samples\_per\_peak})$, where $T$ is the time baseline of each light curve. We tested $\texttt{samples\_per\_peak}=5$, $10$, and $20$. Considering only stars with physical $Q$ values for both periods, we found $36\,790$, $36\,750$, and $36\,759$ sources, respectively, for which $Q(P_1)$ and $Q(P_2)$ were measured. Among these, the strongest Lomb--Scargle peak gave the minimum $Q$ period in $67.9\%$, $67.6\%$, and $67.7\%$ of cases for $\texttt{samples\_per\_peak}=5$, $10$, and $20$, respectively, which is consistent with the $67.9\%$ obtained with the fixed grid. This shows that the need to compare folded light curve quality across multiple independent periods is not due to insufficient frequency sampling.\\

\subsection{Periodic light curves: LPV and YSO candidates}

After validating our period and $Q$ selection in the $Q$ against $M$ space, we filtered out sources showing large asymmetries or aperiodic behavior, yielding the following selection function: $0 \leq Q\leq0.6$ and $-0.4\leq M \leq 0.4$. We classify these as periodic and symmetric light curves in the VVVX data based on their measured $Q$ and $M$ parameters \citep{OrdenesHuanca_2022}. This selection function yielded $15\,010$ periodic and symmetric light curves. From all characteristics computed for the light curves (namely, $Q$, $M$, amplitude, and period), we established different selection functions to identify possible LPVs and YSOs. First, since LPVs typically show $K_\mathrm{s}$ amplitudes higher than $0.4\,\mathrm{mag}$ and periods larger than $80\,\mathrm{days}$ \citep{Matsunaga_2009, Nikzat_2022}, we filtered the periodic and symmetric light curves considering raw amplitudes of $\Delta K_\mathrm{s}=K_{\rm{s,max}}-K_{\rm{s,min}}\geq 0.35$ and $P(K_\mathrm{s})\geq 80\,\rm days$, resulting in a sample of $7616$ stars. The full selection steps for the LPV sample, from the initial VIRAC2$\times$SPICY match down to the final catalog, are summarized in Table~\ref{tab:period_search_LPV}. The amplitude limit considered here is slightly below the one mentioned in the literature, and this is motivated by the fact that the errors in the photometric measurements could affect the measured amplitude.\\

On the other hand, to look for spot-related variations in YSO candidates, we considered the periodic and symmetric light curves that show amplitudes $\Delta K_\mathrm{s} <0.4\,\mathrm{mag}$ and period values $\leq 50\,\mathrm{days}$. Under these criteria, $2004$ sources met these conditions, representing only a small fraction of the entire dataset. The selection steps for the final YSO sample are summarized in Table~\ref{tab:period_search_YSO}. According to the literature, stars in this evolutionary stage are expected to show periodic variations with amplitudes of $\sim 0.2\,\rm mag$ \citep{Wolk_2013, Fischer_2023}. However, we relaxed the amplitude limit for YSO candidate selection because photometric errors affect light curve epochs and because the definition of amplitude varies across studies.\\

We emphasize that the $\Delta K_\mathrm{s}\sim0.35\,\rm mag$ threshold for possible LPVs is not intended to define a physical boundary between YSOs and LPVs and is, instead, motivated by near-IR studies of Mira-like AGB/LPV stars, for which $K$- or $K_\mathrm{s}$-band amplitudes larger than $\sim0.4\,\rm mag$ are commonly used to identify high-amplitude pulsators and by the need to reduce contamination from the overlap region shared by YSOs and LPVs in the amplitude regime. Lower-amplitude LPVs, particularly semi-regular variables, can fall below the imposed amplitude threshold. At the same time, YSOs can exceed it due to accretion or disk-related variability, which is often aperiodic. The period-amplitude ranges occupied by YSOs and semi-regular variables in the literature overlap, particularly in the range of $20\leq P\leq 50\,\rm days$ and $\Delta K_{\rm{s}}\lesssim 0.4\,\rm mag$. Therefore, we should consider sources in this overlap region carefully.\\

One possible reason why we found a small fraction of likely periodic YSOs is that the SPICY catalog is dominated by very red young stars that still retain substantial envelopes or disks. In those cases, accretion drives the variability, making spot-like variations more difficult to detect. The $Q$ metric also depends on the photometric uncertainties. When these are sufficiently large relative to the intrinsic scatter of the light curve, the periodicity metric can formally yield non-physical values below or above its physical limit ($0\leq Q \leq 1.0$). This occurs preferentially for very faint stars and short periods, a regime in which highly reddened YSOs are common. Rather than attempting to retain individual sources with non-physical $Q$ values, we adopted a conservative approach and excluded any source for which $Q(P_1)$ or $Q(P_2)$ was not physically defined ($0<Q<1$).\\

Nevertheless, to assess whether our YSO selection introduces observational biases, we compared the $2004$ automatically selected short-period, low-amplitude (associated with possible YSOs), symmetric candidates with the rejected sources from the same symmetric light curve sample. The selected sources are slightly fainter, with median $K_\mathrm{s}=13.77\,\rm mag$ compared with $K_\mathrm{s}=13.71\,\rm mag$ for the rejected sample, and share similar median photometric uncertainties of about $0.024$ and $0.022\,\rm mag$, respectively. The biggest difference, however, is that the selected sample has a median $\Delta K_\mathrm{s}/\sigma_{K_\mathrm{s}}=9.55$, while the rejected sample has a median value of $18.25$. This is expected because the YSO selection explicitly favors low-amplitude variables, whereas the rejected sample includes higher-amplitude sources, such as possible LPVs. Therefore, the small selected fraction should not be interpreted as a physical lack of spot-modulated YSOs; it reflects the observational and methodological selection imposed by the amplitude condition.

\subsection{Visual inspection of the periodic candidates}
\label{sec:visual_inspection}

Next, we analyze the joint catalog resulting from applying either criterion mentioned in the previous section. To be conservative, we first excluded LPV candidates with unreliable period or amplitude measurements before visual inspection (see Sect.~\ref{sec:period_search}): periods within $0.5\,\rm days$ of the search boundary ($1333.33\,\rm days$), fewer than $100$ raw $K_\mathrm{s}$ epochs, or robust amplitude $\Delta K_\mathrm{s} = K_{\mathrm{s},95} - K_{\mathrm{s},5} < 0.35\,\rm mag$. This reduced the LPV sample from $7616$ to $2785$ sources; we applied no equivalent pre-selection to the $2004$ YSO candidates, all of which we visually inspected. A single inspector examined the raw and phase-folded $K_{\rm s}$ light curves of each candidate, accepting sources with a coherent, low-scatter phase-folded pattern; ambiguous cases were re-folded on the five strongest Lomb--Scargle peaks and retained, with an updated period, if any alternative fold appeared periodic. We note the lack of a formal inter-rater reliability assessment as a limitation of the present catalog.\\

In addition to the main LPV sample described above, we also inspected sources outside the symmetric $-0.4 \leq M \leq 0.4$ box, corresponding to bursting or dipping behavior, since LPVs can show secular brightening or fading over timescales of years \citep{Albarracin_2025} that shifts them outside the symmetric asymmetry range even when the underlying variability is genuinely periodic. We defined this burster/dipper candidate sample using the same period ($P \geq 80\,\rm days$) and periodicity ($0 \leq Q \leq 0.6$) criteria as the main LPV sample. We pre-filtered it with the same automated criteria for symmetric sources (period boundary, $N_\mathrm{epochs} \geq 100$, and robust amplitude $\geq 0.35\,\rm mag$), yielding $2800$ candidates. As a conservative choice, we visually inspected only those with $Q \leq 0.3$ ($851$ sources) and proposed $183$ of them as genuine periodic LPVs. We added these to the main LPV sample.\\

We note that visual inspection is a common final step in light curve morphology studies that use the $Q$ and $M$ metrics and to validate the automated classification (\citeauthor{Cody_2014} \citeyear{Cody_2014}, \citeauthor{Hillenbrand_2022} \citeyear{Hillenbrand_2022}, \citeauthor{Mas_2026} \citeyear{Mas_2026}). These metrics quantify periodicity and flux asymmetry and guide classification of young star variability classes. Still, they do not capture all features of real survey light curves, such as aliases, outliers, poor phase coverage, or mixed variability mechanisms, and they can show discrepancies between measured values and what is actually observed in a light curve, particularly in the imposed boundaries for each class. Therefore, following previous studies based on the $Q$ and $M$ parameters, we used automatic period, amplitude, $Q$, and $M$ cuts to define a reduced candidate set, then visually inspected the raw and phase-folded light curves as a final validation step. This stage is not intended as a scalable classification method for future surveys, but as a conservative validation step for the present sample.\\

For sources whose adopted period did not visually produce a coherent phase-folded light curve, we recomputed the period from an independent, fine-frequency Lomb--Scargle search. Additionally, for sources whose visually preferred period differed from the one that automatically minimized $Q$ and was $\geq 80$ days, we tested local refinement (including half- and double-period windows, as described in Sect.~\ref{sec:period_search}). For each final adopted period, we computed the false-alarm probability (FAP) by evaluating the Lomb--Scargle power at the corresponding frequency and using the analytic Baluev method implemented in \texttt{astropy.timeseries.LombScargle}. We therefore interpret our final proposed LPV and YSO lists as high-confidence, conservative selections rather than complete catalogs. A total of $16$ LPV candidates remained ambiguous after this process and were conservatively excluded from the final catalog. To make the procedure reproducible, we provide the final tables with the adopted period, amplitude, $Q$, $M$, and FAP (where available), for each probable stellar type.\\

The complete visual inspection revealed that $742$ LPV candidates exhibit periodic, mostly sinusoidal variations. Their photometric properties, along with periods, raw $K_\mathrm{s}$ amplitudes ($\Delta K_{\rm{s, max}}-K_{\rm{s, min}}$), $Q$ and $M$ metrics, are listed in Table~\ref{tab:lpv_sample}. In all visually checked LPVs, $741/742$ ($99.9\%$) have periods with FAP $< 10^{-3}$, further supporting their periodicity. Among the selected YSOs, $307$ out of $2004$ sources display periodic patterns in our visual inspection and are listed in Table~\ref{YSOs_table}. Nevertheless, $245$ have periods with FAP $< 10^{-3}$, corresponding to $79.8\%$ of our YSO sample. Even though the remaining sources passed our visual inspection for periodic patterns, their period values need further confirmation. FAP values for YSOs are listed in Table~\ref{YSOs_table}.\\

\begin{figure}
        \centering
        \includegraphics[width=0.9\hsize]{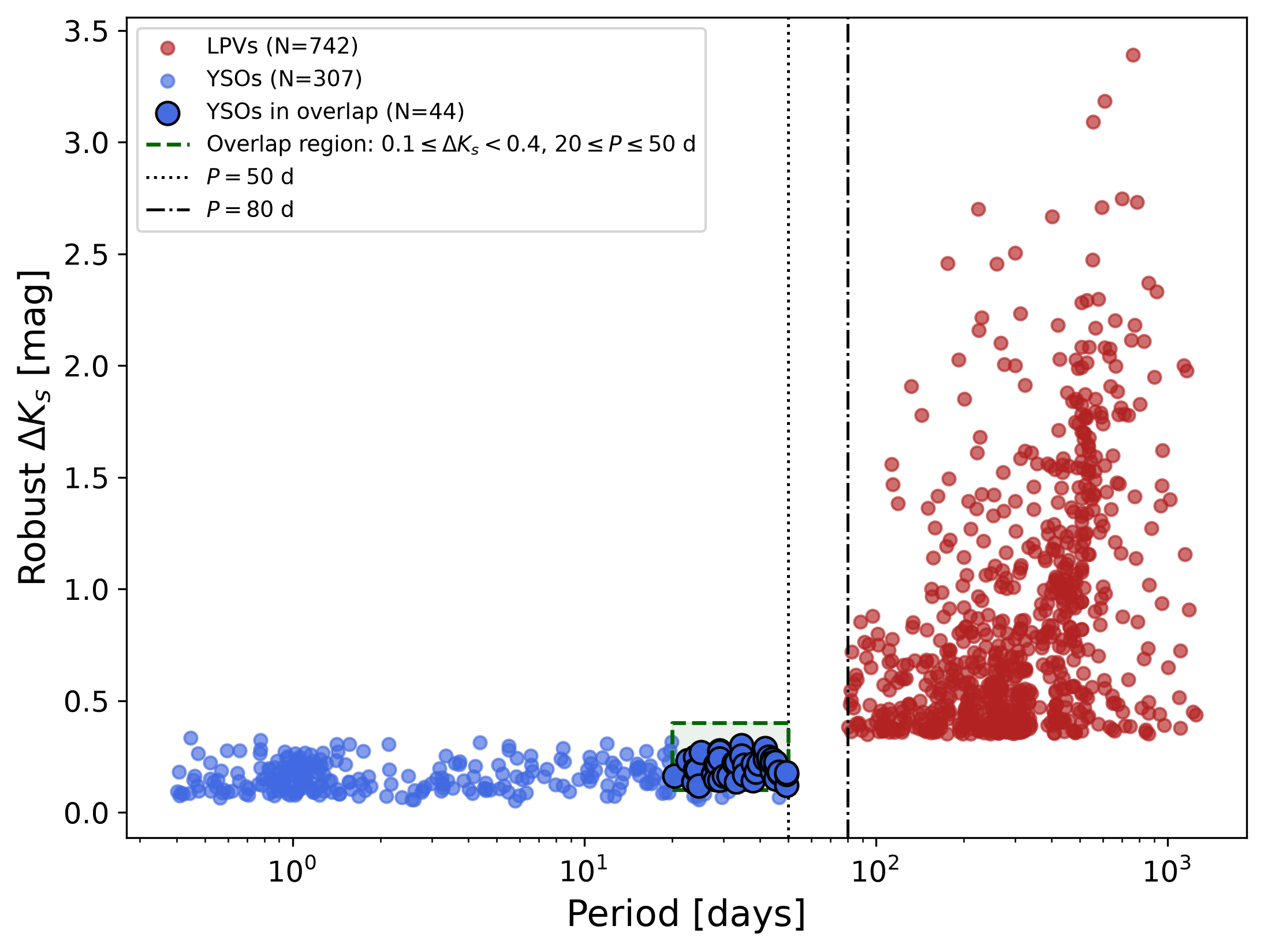}
        \caption{Robust amplitude $\Delta K_{\rm{s}}= K_{\rm{s},95}-K_{\rm{s},5}$ versus the computed period for our visually selected YSOs and LPVs. The overlap region between semi-regular evolved variables and YSOs is marked by a green-shaded rectangle and contains $44$ of our selected YSOs.}
        \label{amp_per}%
\end{figure}

To analyze the overlap region between YSOs and semi-regular variables, we computed a robust amplitude for all our visually selected sources, considering the $95$th and $5$th percentiles of their $K_\mathrm{s}$-band, robust $\Delta K_\mathrm{s} = K_{\mathrm{s},95} - K_{\mathrm{s},5}$. Our YSO selection contains $44$ stars ($\approx 14.3\%$) of this sample that are in the overlap period and amplitude region, which they could share with evolved semi-regular variables. These stars are highlighted in the robust $\Delta K_{\rm{s}}$ amplitude against period plot shown in Fig.~\ref{amp_per}. This plot also shows a tendency for LPVs with higher amplitudes to have longer periods, which is expected for Miras and semi-regular variables \citep{Soszynski_2011, Iwanek_2021} and further supports our selection.\\

\begin{figure*}[ht!]
        \centering
        \includegraphics[width=0.8\hsize]{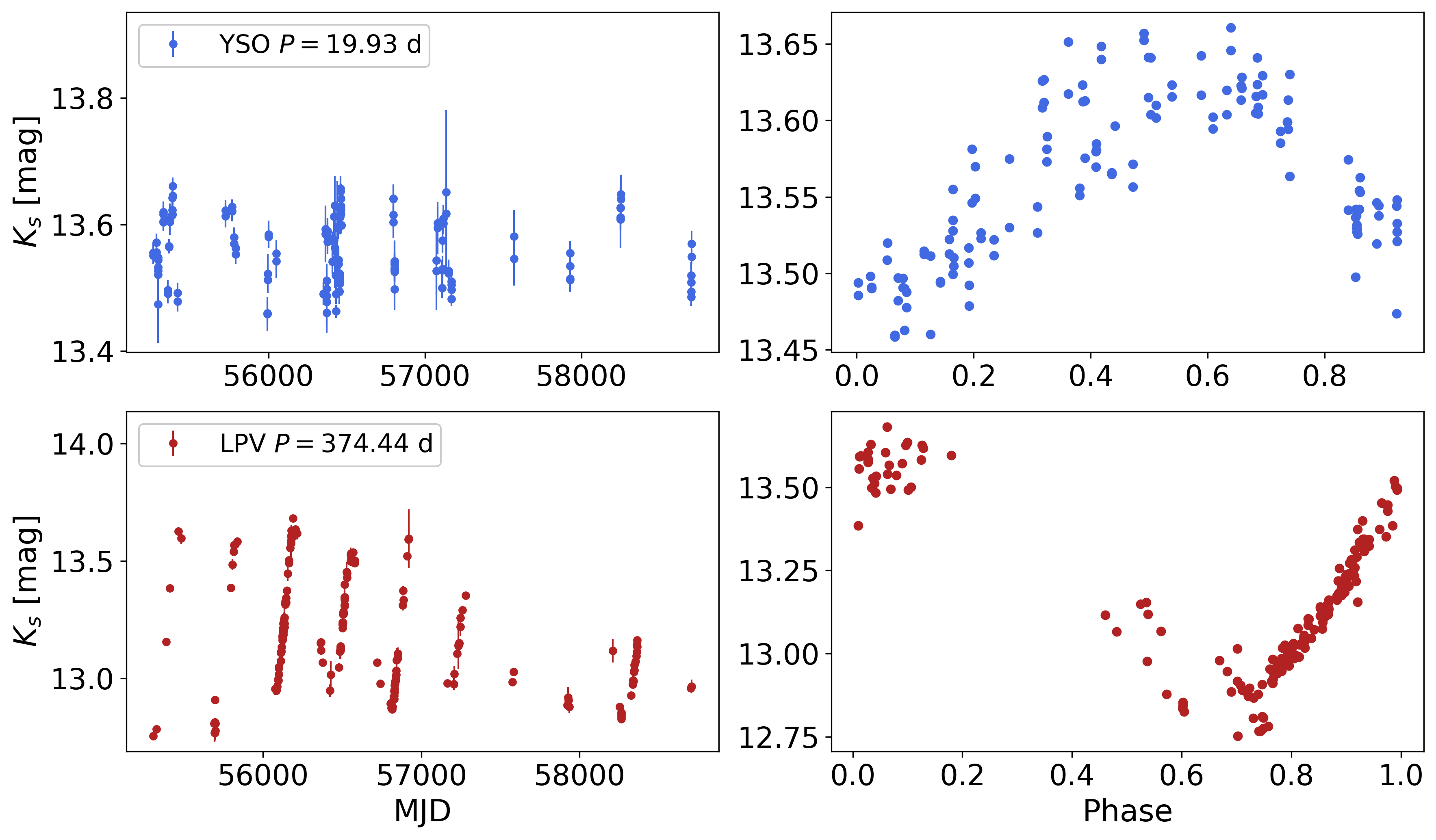}
        \caption{Example light curves of a source classified as a YSO (upper panels) and a source classified as an LPV (lower panels). Raw light curves are shown on the left, whereas phase-folded ones, according to the indicated period $P$, are on the right.}
        \label{examples}%
\end{figure*}

Light curve examples are shown in Fig.~\ref{examples}. The upper panels show the raw (left) and phase-folded light curve (right) of one YSO in our sample (both in blue). The lower panels show the same for an LPV star (both in dark red). The derived periods are indicated in each legend. The LPV's almost sinusoidal variation is already observed in its raw light curve, particularly between $56\,000 \leq$ MJD $\leq 57\,000$. In contrast, the periodic variation of the YSO becomes apparent only after phase folding.

\subsection{Spatial distribution}

\begin{figure*}
        \centering
        \includegraphics[width=0.75\hsize]{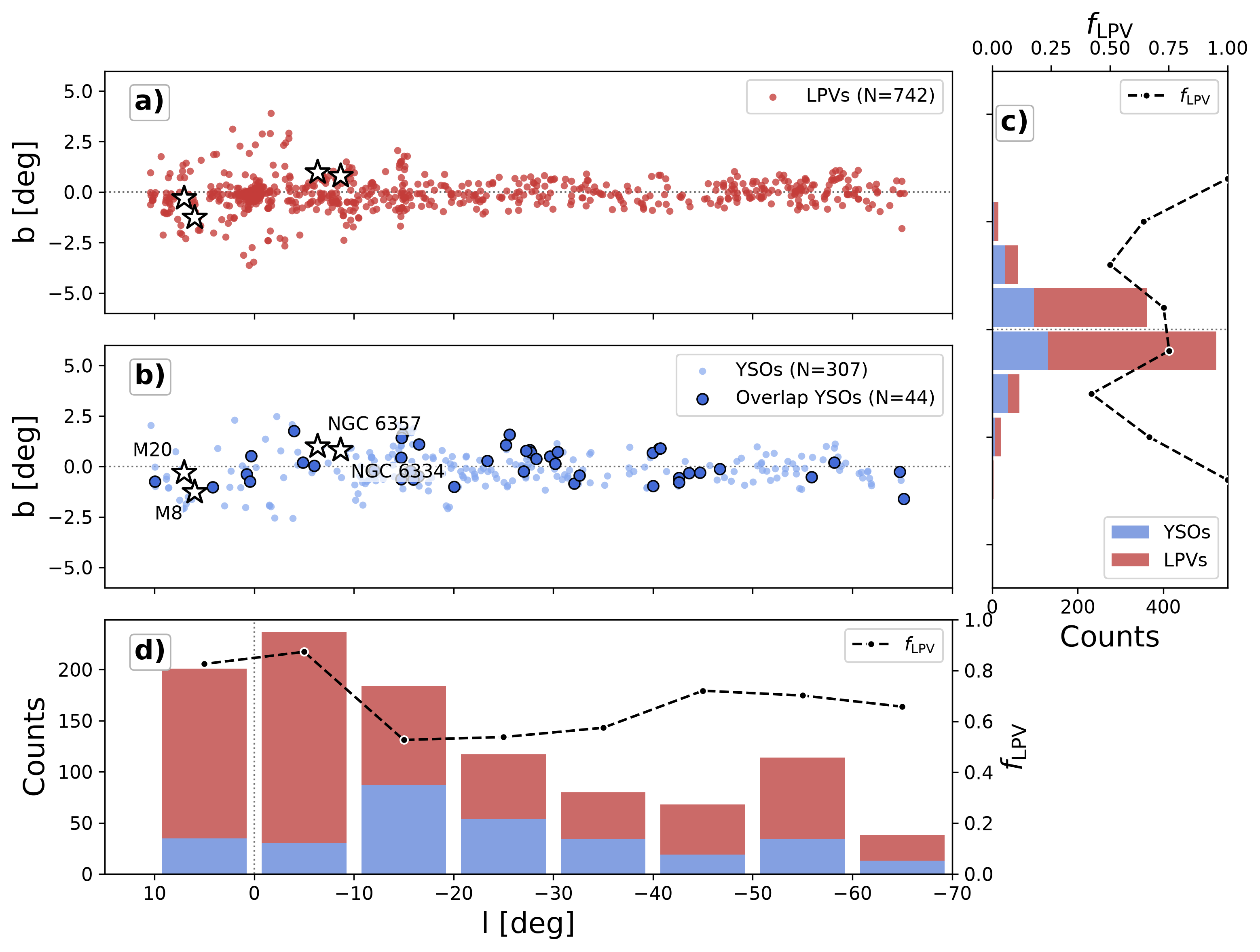}
        \caption{\textit{Panel a):} Spatial distribution of our selected LPVs (red dots) in Galactic coordinates. Black star symbols mark the Galactic coordinates of the massive star-forming regions discussed in the text. \textit{Panel b):} Same as panel a), but now for our list of YSOs (blue shaded dots). Those in the overlap region are marked in darker blue and encircled in black. \textit{Panel c):} Histogram of the Galactic latitude $b$ distribution of LPVs and YSOs, following the same color-coding as the other panels and binned in $1\degree$ width. The fraction of LPVs, $f_{\rm{LPV}}$, is shown on the upper horizontal axis, and the raw counts are on the lower horizontal axis. \textit{Panel d):} Same as panel c), but for the Galactic longitude distribution. The fraction of LPVs, $f_{\rm{LPV}}$, is in bins of $10\degree$.}
        \label{spatial}%
\end{figure*}

\begin{figure}[ht!]
        \centering
        \includegraphics[width=\hsize]{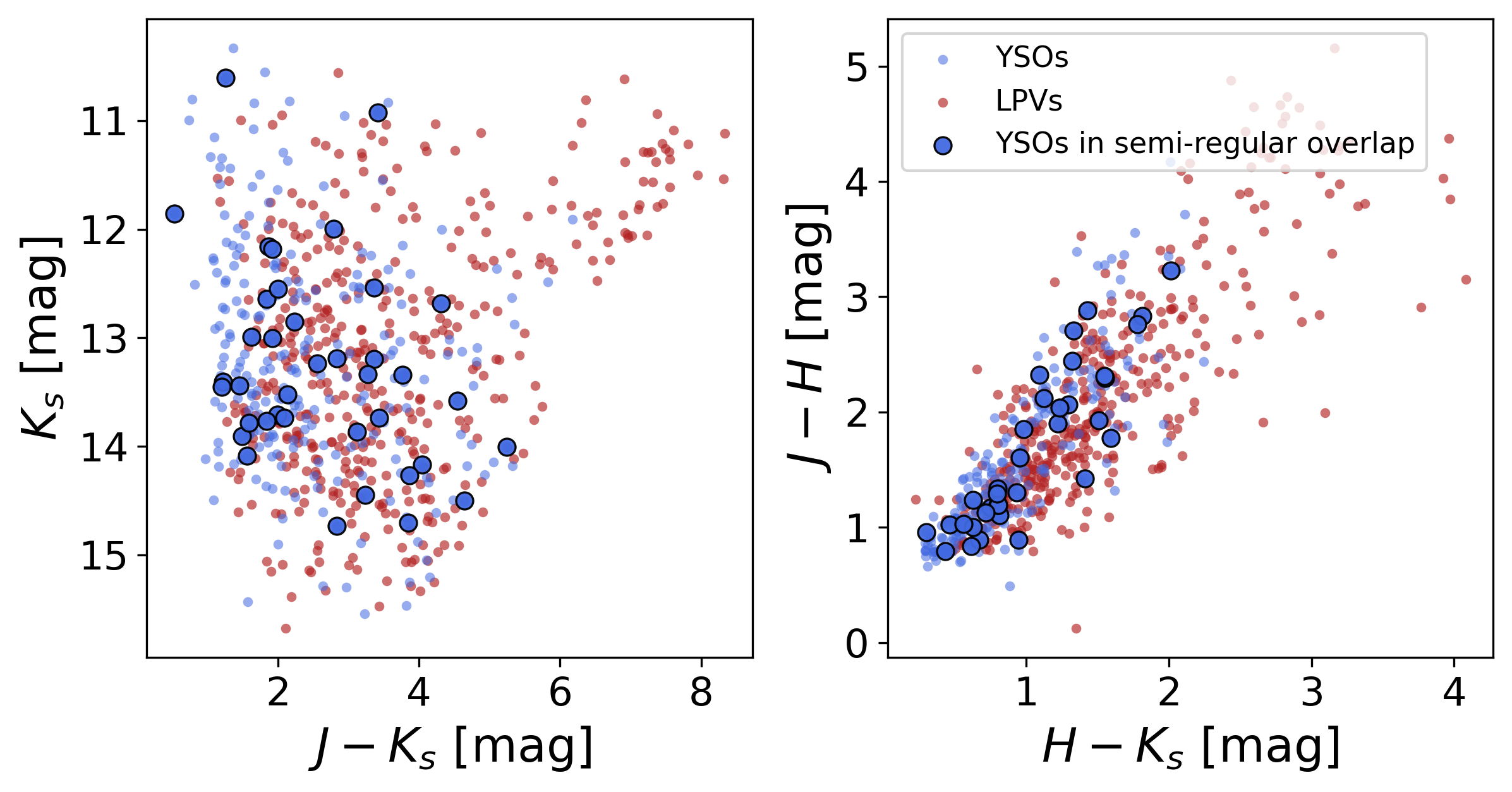}
        \caption{\textit{Left panel:} $K_\mathrm{s}$ against $J-K_\mathrm{s}$ CMD of LPVs and YSOs selected by visually inspecting their light curves. \textit{Right panel:} $J-H$ versus $H-K_\mathrm{s}$ color-color diagram for the same sources. YSOs encircled in black are the ones that can have periods and amplitudes similar to semi-regular variables.}
        \label{cmd_ccd}%
\end{figure}

The spatial distributions of the two populations are shown in Fig.~\ref{spatial}, together with the locations of $4$ massive star-forming regions (white stars), namely M20, M8, NGC~6357, and NGC~6334. LPVs (upper panel) are highly concentrated towards the Bulge, whereas YSOs (central panel) are predominantly distributed along the Milky Way plane. The lower panel in Fig.~\ref{spatial} shows a histogram for the fraction of LPV stars $f_{\rm{LPV}}=N_{\rm{LPV}}/(N_{\rm{LPV}}+N_{\rm{YSO}})$ among the total number of LPVs and YSOs, in bins of$10$ degrees of Galactic longitude and highlights that they outnumber the YSOs particularly towards the Bulge ($|l|\leq 10\degree$). YSOs in the overlap region between them and semi-regular variables are shown with black circles. Most are located in the MW plane and toward its disk. Some of these YSOs are within the field of view of the Bulge, and several are near a star-forming region, but others lie outside the Galactic plane. These should be treated with caution. However, the overall distribution of our selection of LPVs and YSOs suggests that star-forming regions towards the Bulge of the MW and any star formation rate that considers young star counts in the same field-of-view of the Bulge of the Galaxy can be severely affected by contamination of LPV stars. Photometry alone is therefore insufficient to disentangle LPV contamination within the population of YSOs. This can be observed in Fig.~\ref{cmd_ccd}, where the two sub-samples overlap in both the CMD and the color-color diagram. For YSOs with periods and amplitudes similar to semi-regular variables (encircled in black), one can observe from the color--color diagram that they do not reach high $J-H$ colors that are more associated with LPVs (upper right part of this plot).

\subsection{Additional near- and mid-IR diagnostics}

\begin{figure}
        \centering
        \includegraphics[width=0.95\hsize]{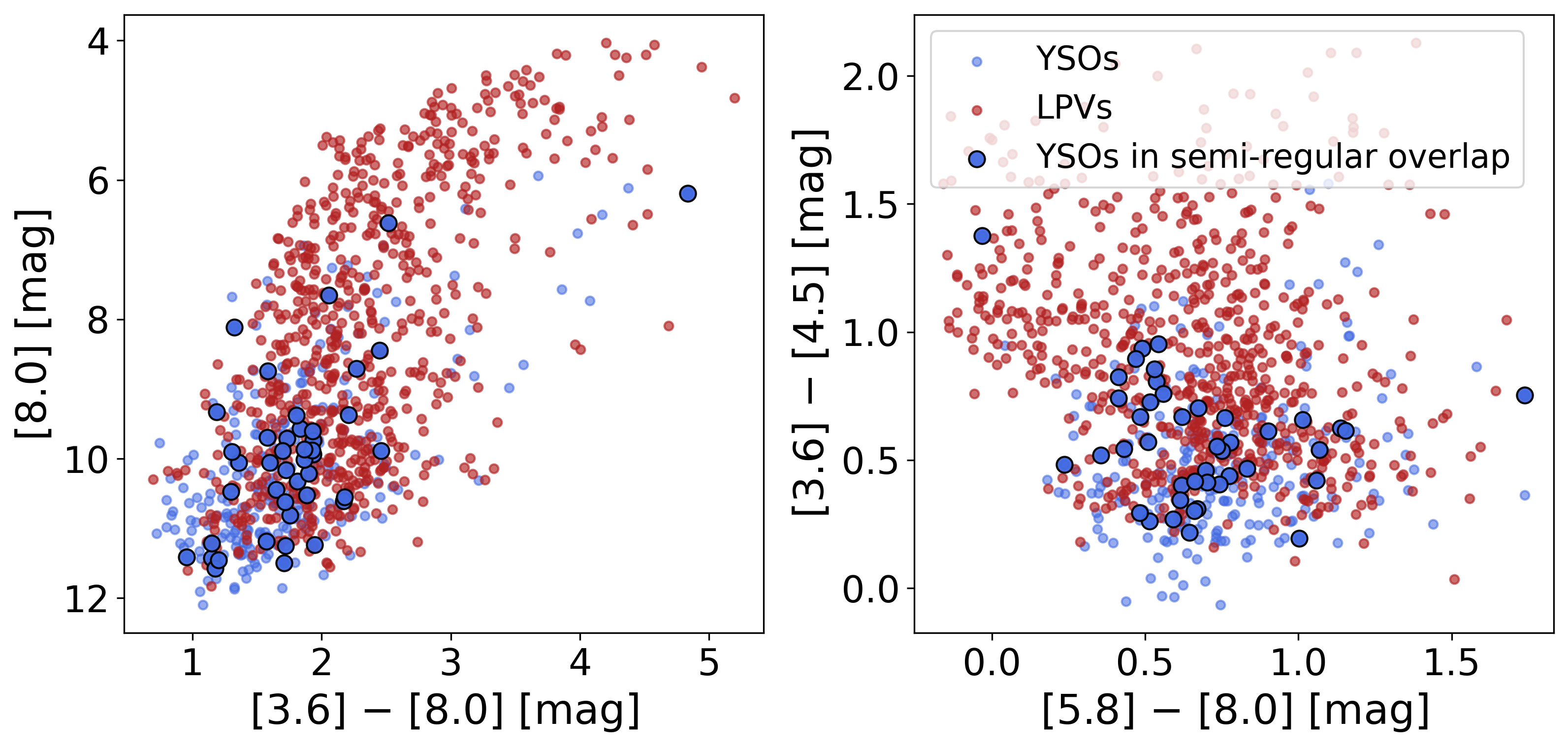}
        \caption{\textit{Left panel:} \textit{Spitzer} $[8.0]$ against $[3.6]-[8.0]$ photometry CMD of our list of LPVs (in dark red) and YSOs (in blue) candidates. \textit{Right panel:} $[3.6]-[4.5]$ against $[5.8]-[8.0]$ color-color diagram for the same sample.}
        \label{cmd_spitzer}%
\end{figure}

IR observations remain an important tool for detecting these highly reddened sources, which could continue hidden at other wavelengths. In particular, mid-IR is sensitive to warm dust from the inner regions of YSO disks, providing valuable diagnostics of their circumstellar environments. \textit{Spitzer} provides random-phase, single-epoch magnitudes for both LPVs and YSO candidates. The $[8.0]$ versus $[3.6]-[8.0]$ CMD is shown in Fig.~\ref{cmd_spitzer} and reveals that our proposed periodic YSOs (shown in blue) cluster at fainter $[8.0]$ magnitudes, including the majority of those in the semi-regular overlap (encircled in black). This may indicate their true nature as young stars. However, mean magnitudes are needed to provide a robust constraint on their CMD position.\\

If pulsations drive the variability in both the mid- and near-IR bands, the corresponding light curves are expected to be correlated \citep{Albarracin_2025}. WISE \citep{WISE_2010} and NEOWISE \citep{NEOWISE_2011} surveys allowed to obtain light curves in $W1=3.4\,\mathrm{\mu m}$ and $W2=4.6\,\mathrm{\mu m}$ passbands. We cross-matched our lists of sources with the WISE/NEOWISE catalogs, using a matching radius of $5''$ to account for the different point spread functions of the instruments and given the WISE angular resolution of $6.1\,\rm arcsec$ and $6.4\,\rm arcsec$ in $W1$ and $W2$, respectively \citep{WISE_2010}. We found that $501$ of our selected LPVs and $154$ of our YSO candidates have at least one $W2$ epoch in the WISE/NEOWISE catalogs. We therefore examined the spread of the data points for each star in the $W2$ band, using WISE/NEOWISE data free of artifacts and excluding upper limits and non-detections. We further restricted the sample to exposures with valid frame quality and S/N $> 5$ per epoch in $W2$, and rejected measurements affected by more than two blended neighbors. Finally, we considered only detections within $2\,\rm arcsec$ of the target position. Sources with at least $10$ epochs surviving these cuts were considered to compute $\Delta W2=W2_{95th}-W2_{5th}$, defined as the difference between the $95$th and $5$th percentiles of the cleaned light curves ($417$ LPV and $120$ YSO candidates).\\

\begin{figure*}
        \centering
        \includegraphics[width=0.85\hsize]{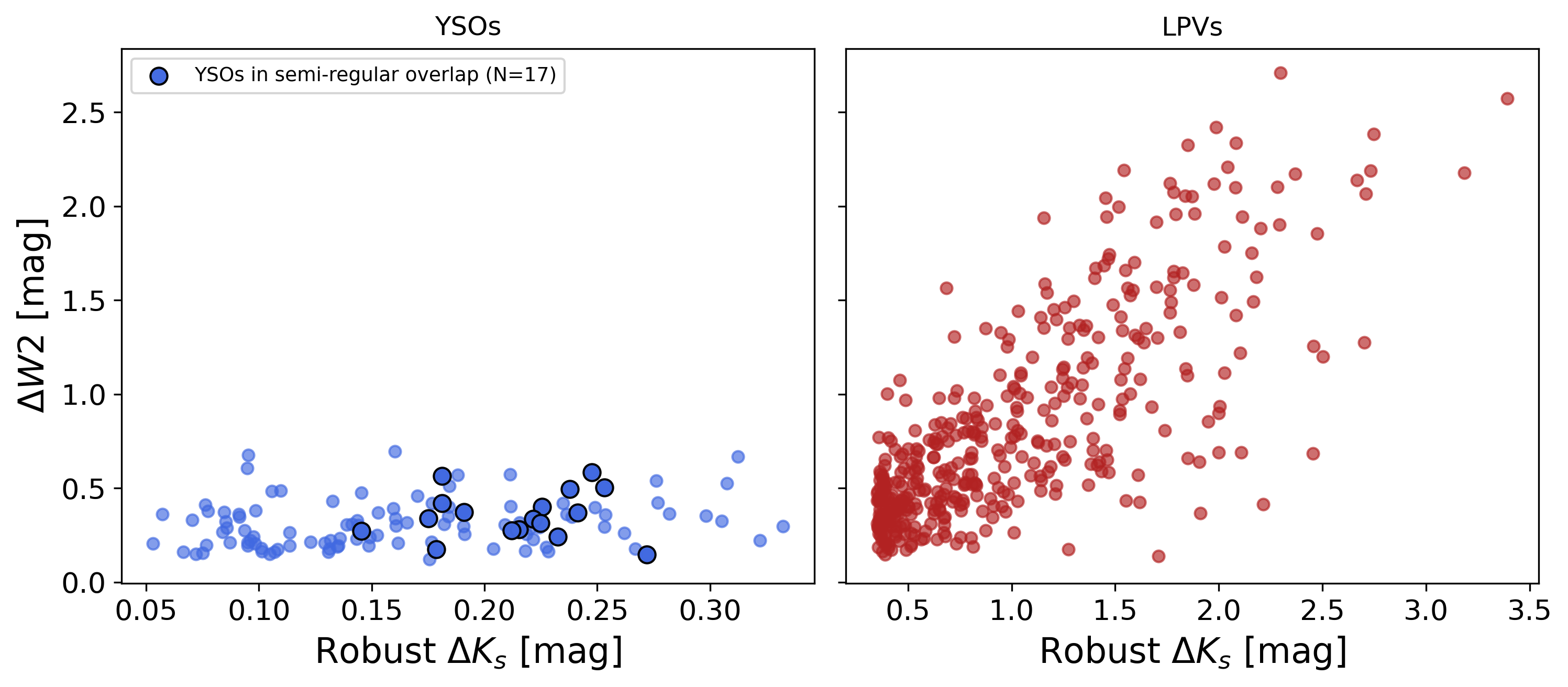}
        \caption{Difference between the $95$th and $5$th percentiles of the cleaned $W2$ light curves, $\Delta W2$, against the $95$th and $5$th percentiles of the $K_\mathrm{s}$-band for our YSO sample (left panel) and LPVs (right panel), Robust $\Delta K_\mathrm{s}$. For LPVs, both amplitudes show a strong correlation confirmed by a Spearman coefficient of $\rho = 0.741$ ($N=417$), in contrast to the weak correlation found for YSOs ($\rho = 0.264$, $N=120$).}
        \label{w2_ks}%
\end{figure*}

In Fig.~\ref{w2_ks}, $\Delta W2$ is plotted against a robust variability amplitude in the $K_\mathrm{s}$-band defined accordingly to the one defined for the $W2$-band as the difference between the $95$th and $5$th percentiles (called robust $\Delta K_\mathrm{s}$). The two quantities show a strong correlation for the LPV sample (right panel), with a Spearman rank correlation coefficient of $\rho=0.741$ ($N=417$, $p=9\times10^{-74}$, bootstrap $95\%$ confidence interval (CI) $0.69$--$0.79$). This indicates that the LPV's observed variability in both bands has a common origin. For YSOs, mid-IR emission is expected to arise mainly from the intermediate regions of their disks \citep{Venuti_2021}. Mid-IR variations can be linked to periodic physical phenomena, such as disk occultation \citep{MoralesCalderon_2011}. However, analyzing the latter is beyond the scope of this study and will be addressed in a forthcoming work. Dusty structures can be the origin of the $W2$ band emission observed in the $120$ YSOs of our sample for which $\Delta W2$ could be measured. For these sources $\Delta W2$ correlates only weakly with robust $\Delta K_\mathrm{s}$ ($\rho = 0.264$, $N=120$, $p=3.6\times10^{-3}$, bootstrap $95\%$ CI $0.09$--$0.44$). The left panel of Fig.~\ref{w2_ks} shows this weak correlation and suggests that near-IR and mid-IR variability in YSOs originate from different physical mechanisms; therefore, near-IR variability is likely driven by stellar spots in the photosphere, whereas mid-IR variability may arise from processes in the disk itself. Only $17$ of the $44$ YSOs in the semi-regular overlap region had available $\Delta W2$ data, and they also do not show a correlation between these amplitudes.\\

\begin{figure}
        \centering
        \includegraphics[width=0.8\hsize]{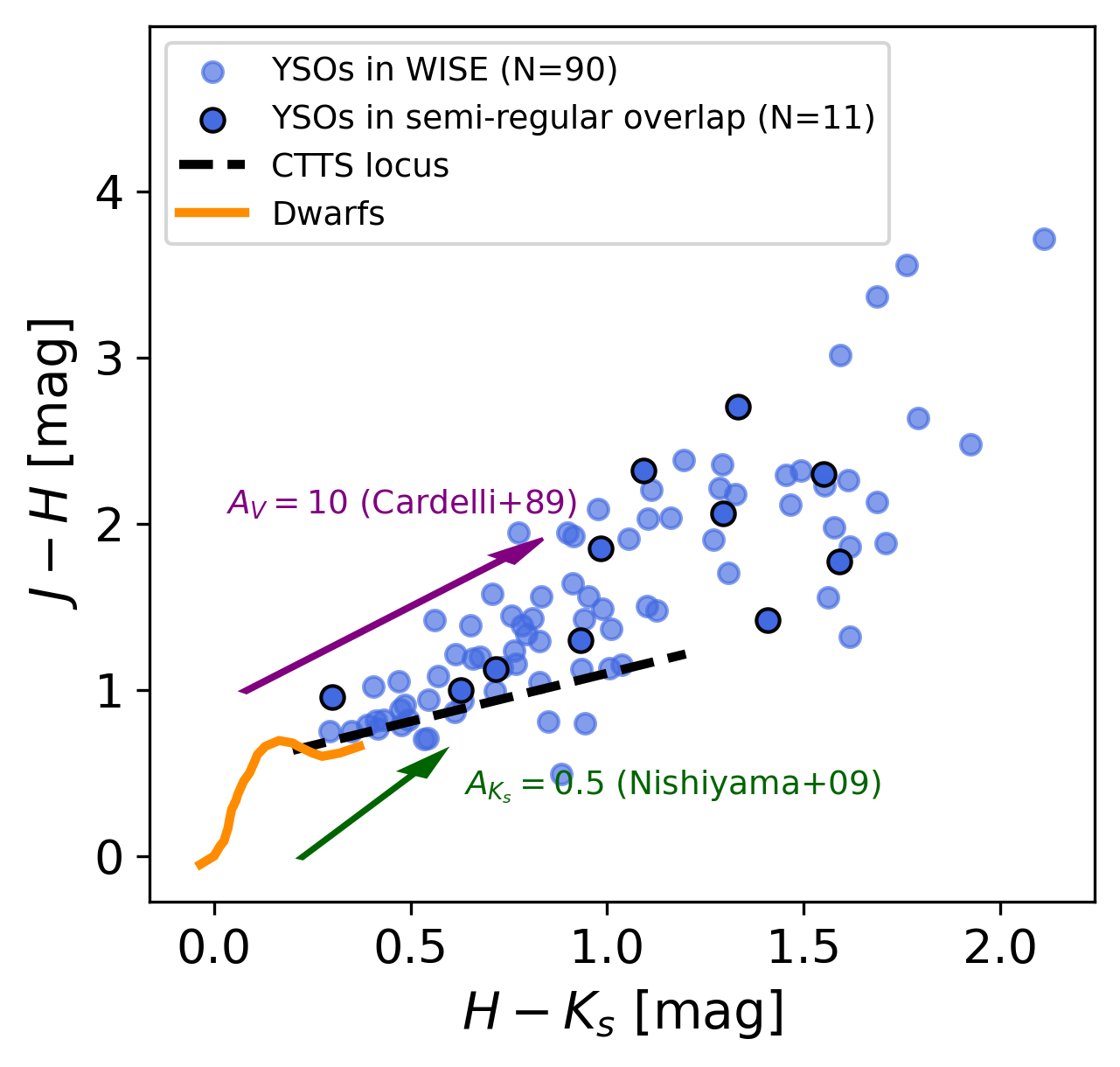}
        \caption{$J-H$ versus $H-K_\mathrm{s}$ color-color diagram for YSOs of our list with WISE $W2$ data with those in the semi--regular period and amplitude overlap encircled in black. The CTTSs locus is depicted as a dashed black line, whereas the location of dwarf stars is shown as a dark orange line. Of the $120$ YSOs with a measured $\Delta W2$ shown in Fig.~\ref{w2_ks}, $90$ ($11$ of them in the overlap region) also have $J$ and $H$ photometry and can therefore be placed in this diagram.}
        \label{ysos_ccd}%
\end{figure}

To confirm whether these sources can host circumstellar disks, we present their near-IR color-color diagram in Fig.~\ref{ysos_ccd}, overlaid with the classical T Tauri stars (CTTSs) locus from \citet{Meyer_1997} as a dashed black line. The orange line shows expected colors of dwarf stellar photospheres \citep{Bessell_1988}. Because these stars can span a wide range of distances and extinctions, we show reddening vectors corresponding to both extinction laws of \citet{Cardelli_1989} and \citet{Nishiyama_2009}. Several sources lie close to the CTTS locus, including $4$ from the semi-regular overlap, suggesting circumstellar disks. However, foreground extinction can shift stars toward the upper-right regions of this diagram, as observed. Even though stars surrounded by disks also develop bright and dark spots on their surfaces, the spot-related variability is often masked by the presence of the disk \citep{OrdenesHuanca_2022}. Fig.~\ref{ysos_ccd} suggests these stars can host circumstellar disks, and they likely contribute to the mid-IR emission and variability amplitudes of these sources, which differ from those observed in the $K_\mathrm{s}$ band. Additionally, the photometric uncertainties of highly reddened sources at the faint end of the VIRAC2 catalog reduce the sensitivity of the $Q$ metric to low-amplitude periodic signals and produce non-physical values beyond its physical limit ($0\leq Q \leq 1$).

\section{Discussion}

\subsection{Comparison with other contamination finding method}

Recently, \citet{Jheonn_2026} applied a machine learning-based classification method to disentangle these two stellar groups. Their approach used time-series data from WISE/NEOWISE measurements, which have a lower cadence than the VVVX/VIRAC2 catalog. In their study, they showed that machine learning techniques can distinguish these two populations, provided they include time-series data. This finding motivates our simpler approach, which emphasizes that photometry alone is insufficient for this task and that time-domain information is essential for a reliable classification.\\

\begin{figure}
        \centering
        \includegraphics[width=0.8\hsize]{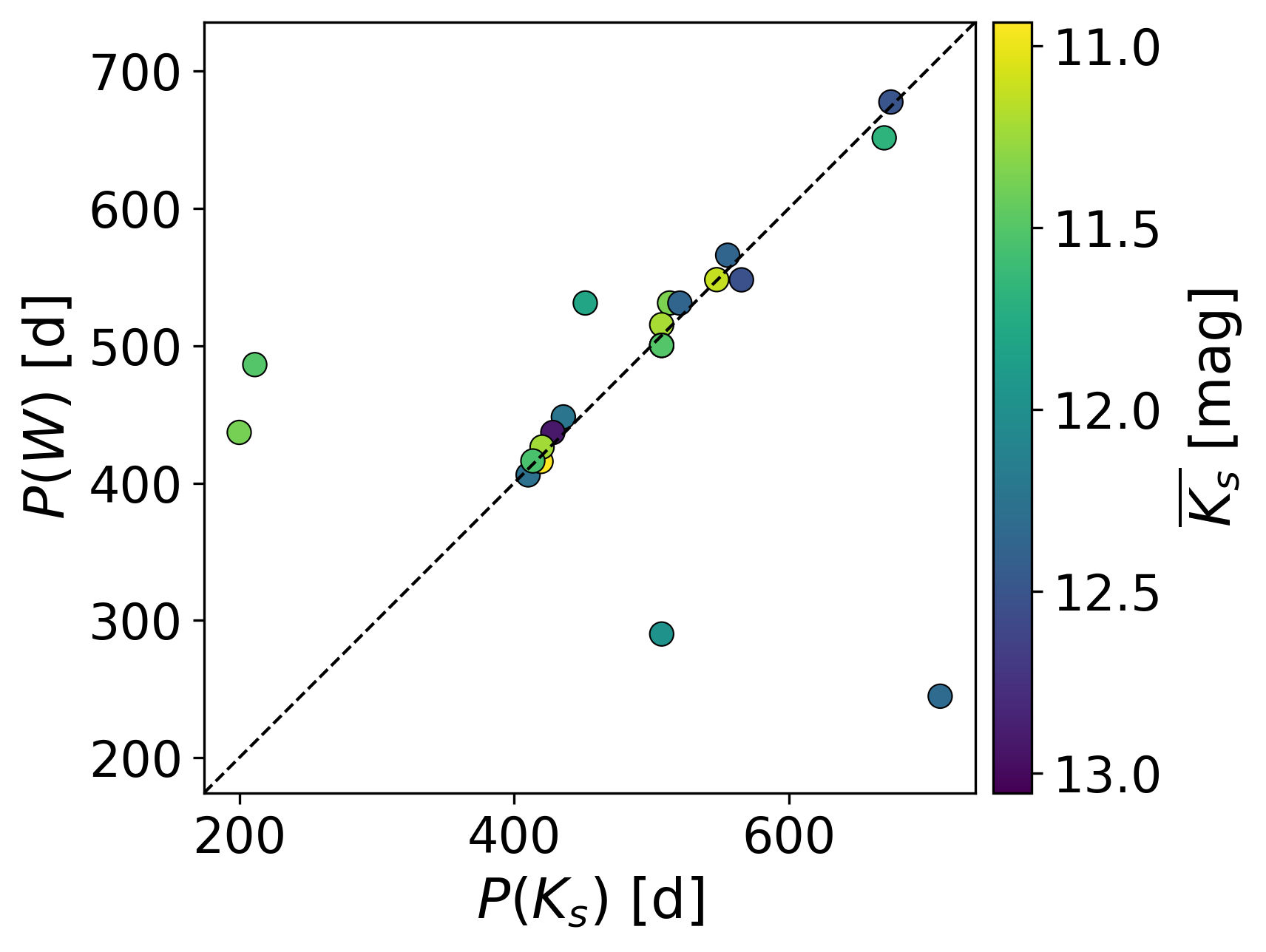}
        \caption{Comparison between the periods derived in this work, $P(K_\mathrm{s})$, and those reported by \citet{Jheonn_2026}, $P(W)$, for the sources in common between both samples. The one-to-one relation is shown as a dashed black line, and the color bar represents the $K_\mathrm{s}$ mean magnitude from VIRAC2.}
        \label{jheonn}%
\end{figure}

By cross-matching our LPV sample with their list of $258$ AGB candidates with a $6\,\rm arcsec$ radius, we found $22$ sources in common. Each has only one match within this region, and all $22$ were marked as variable in their dataset. The periods obtained in this work $P(K_\mathrm{s})$ compared to theirs $P(W)$ are shown in Fig.~\ref{jheonn} and exhibit good agreement, with a median difference of $\sim 9.5\,\rm days$ and a median fractional difference $|P(K_\mathrm{s}) - P(W)|/\langle P \rangle$ of $0.019$ (with $\langle P \rangle$ being the mean period $(P(K_\mathrm{s}) + P(W))/2$). However, there are $4$ common variable sources that have period values that do not agree. Since the color bar in Fig.~\ref{jheonn} shows their mean $K_{\rm{s}}$ magnitude, this discrepancy cannot be attributed to their faintness. For these $4$ discrepant sources, we also checked their top three most probable periods from an independent fine-frequency Lomb--Scargle search. For $3$ of the $4$ sources, one of these alternative candidate periods lies within $5\%$ of the value reported by \citet{Jheonn_2026}. However, our minimum $Q$ choice did not automatically select this period for these sources. For the remaining source, none of the three most probable periods are close to their reported value. We attribute these discrepancies to differences in sampling, number of epochs, the baselines used in the period computation, and to the minimum $Q$ value selection occasionally preferring an independent peak over the one closest to the WISE-derived period. In addition, none of our selected YSOs appear on their AGB list. These strongly support the reliability of our LPV and YSO identifications.

\subsection{Complementary X-ray and mid-IR information for both populations}

Together, this evidence highlights the value of monitoring stars over many years when examining the separation between the two populations. Because such surveys are observationally expensive in telescope time, X-ray observations provide a powerful complementary tool: they can overcome extinction and are known to be weak or absent in AGB stars \citep{Ortiz_2021}. Thus, we cross-matched our list of AGB contaminants with the \textit{Chandra} Source catalog v2.1 \citep{Weisskopf_2002}, using a $1\,\rm arcsec$ matching radius because of its angular resolution. Only $5$ out of $742$ sources were found to have X-ray emission.\\

On the other hand, from our list of proposed YSOs, only \textcolor{red}{$10$} were also in the \textit{Chandra} catalog. This may relate to the accretion status of these young stars. Accretion events can suppress X-ray emission. Non-accretor sources show higher X-ray activity than accretors \citep{Preibisch_2005}. Since the sources from our YSO list should contain disks or envelopes, their X-ray emission can be low and, thus, not detected by \textit{Chandra}. \citet{Kuhn_2021} also addressed the potential AGB contamination in their YSO sample. For instance, the catalog presented by \citet{Robitaille_2008} contains intrinsically red sources identified by \textit{Spitzer}/GLIMPSE (Galactic Legacy Infrared Mid-Plane Survey Extraordinaire; \citet{Benjamin_2003}) data located in the Galactic mid-plane. When comparing that catalog with SPICY, the authors found that $\sim 35\%$ of their proposed YSOs could be misclassified. Our list of AGB variables comprises only $\approx 1.26 \%$ of the sources in common between VIRAC2 and SPICY analyzed in this study, suggesting a lower limit of contamination.\\

For comparison, we cross-matched our final AGB and YSO samples against the catalog of intrinsically red sources from \citet{Robitaille_2008}, using a $1\,\rm arcsec$ radius (IRAC angular resolution is $\sim 2\,\rm arcsec$). We found $271$ matches among our $742$ LPV candidates and $40$ matches among our YSO sample. The \citet{Robitaille_2008} diagnostic requires valid $[4.5]$, $[8.0]$, and $[24]$ photometry; therefore, the following comparison is restricted to the sources satisfying this requirement, $247$ AGB candidates and $30$ YSO candidates. Using the \citet{Robitaille_2008} AGB criterion, $[4.5]\leq 7.8$ or $[8.0]-[24]<2.5$, $153$ of the $247$ AGB candidates fall in the AGB region, corresponding to an agreement of $61.9\%$. Conversely, using the \citet{Robitaille_2008} YSO criterion, $[4.5]>7.8$ and $[8.0]-[24]\geq2.5$, $21$ of the $30$ YSOs from our list fall in their YSO region, corresponding to an agreement of $70\%$.\\

Robitaille's YSO region also contains $94$ of our final AGB candidates (internal contamination of $81.74\%$) for YSO selection based on this color--magnitude criterion, assuming our classification is correct. Nevertheless, $8$ out of $94$ discrepant sources are classified as Mira by \citet{Albarracin_2025}. In contrast, Robitaille's AGB region is relatively clean, with only $9$ of our final YSOs falling in this region, corresponding to an internal contamination of $\approx5.5\%$. Thus, the color--magnitude regions defined in \citet{Robitaille_2008} recover broad trends but do not cleanly separate the two populations in our sample. This confirms quantitatively that color--magnitude cuts alone are insufficient to separate YSOs from evolved red variables.\\


\begin{figure*}[ht!]
        \centering
        \includegraphics[width=\hsize]{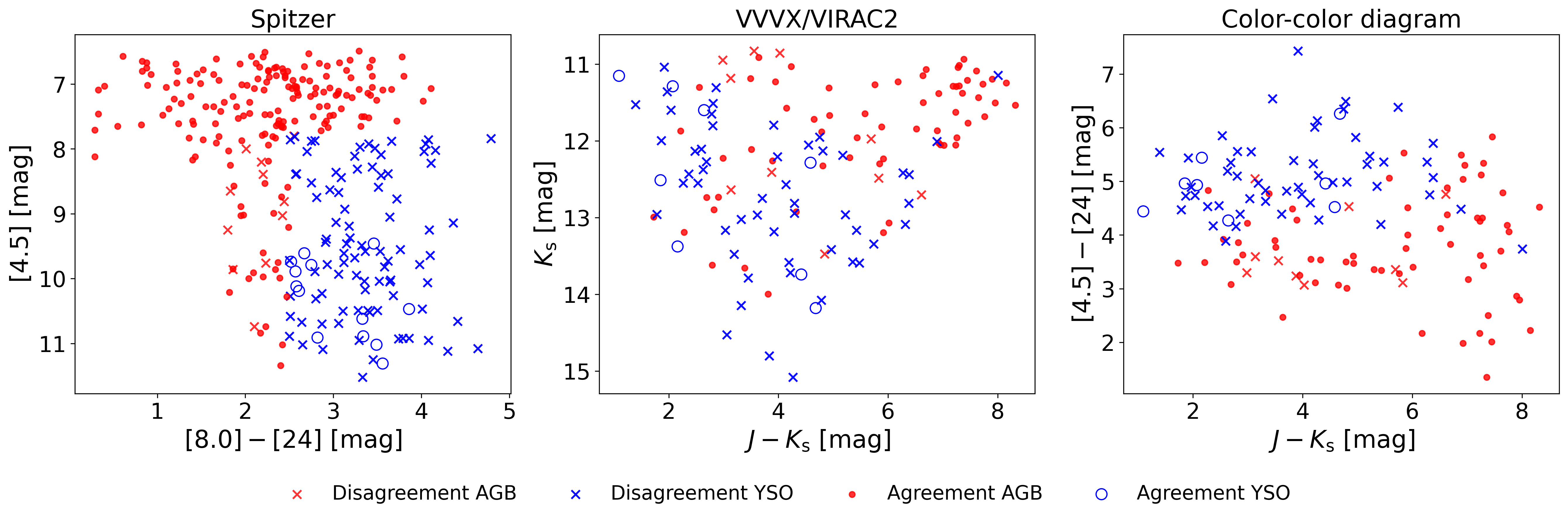}
        \caption{\textit{Left panel:} \textit{Spitzer} $[4.5]$ versus $[8.0]-[24]$ CMD for YSOs and AGB stars in common with the catalog of red sources by \citet{Robitaille_2008}. \textit{Central panel:} VVVX/VIRAC2 $K_\mathrm{s}$ versus $J-K_\mathrm{s}$ CMD for the same sources. \textit{Right panel:} color-color diagram using \textit{Spitzer} $[4.5]-[24]$ and VIRAC2 $J-K_\mathrm{s}$ colors. In all three, red filled circles (AGBs) and open blue circles (YSOs) indicate classifications that agree between their study and ours. Crosses mark discrepancies between both, using the same color code.}
        \label{robitaille}%
\end{figure*}

Fig.~\ref{robitaille} presents the \textit{Spitzer} and VVVX/VIRAC2 CMDs for the sources in common, shown in the left and central panels, respectively. The $[4.5]$ versus $[8.0]-[24]$ diagram corresponds to the magnitude and color criteria adopted in the catalog of red sources to separate AGB stars from YSOs. Red-filled circles (AGBs) and blue open circles (YSOs) indicate sources for which the two studies agree on the classification. The central panel contains fewer objects because some of the common sources lack $J$-band measurements. Despite this limitation, AGB stars tend to occupy redder $J-K_\mathrm{s}$ colors than the small number of YSOs in agreement. A similar trend is seen in the color-color diagram (right panel). However, a substantial overlap between the two populations remains. We stress here that their labels are not fully independent, nor confirmed, and they serve us only as a comparison tool.\\

\begin{figure}[ht!]
        \centering
        \includegraphics[width=0.95\hsize]{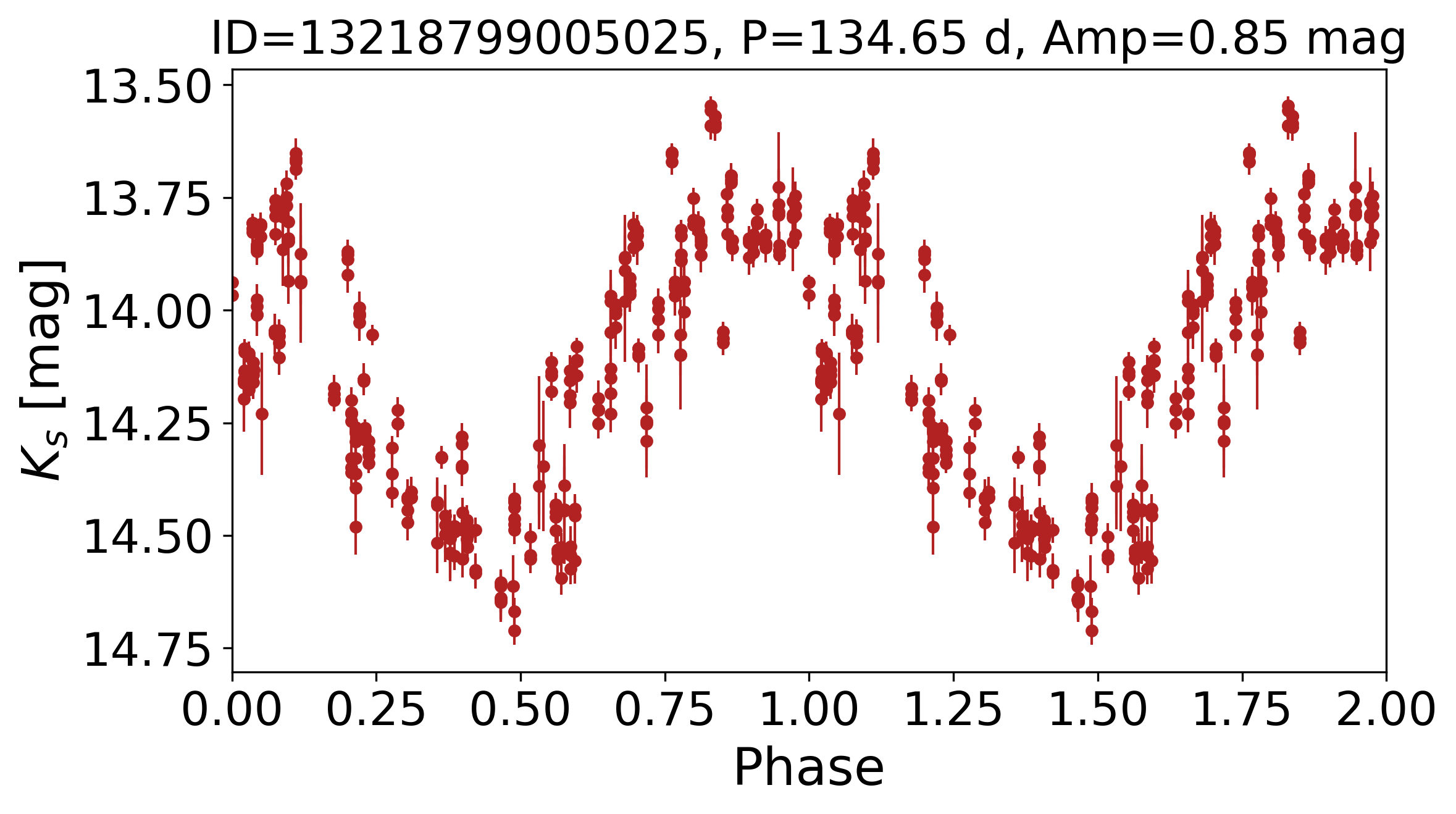}
        \caption{Example of a phase-folded light curve for a star in our AGB/LPV list that did not meet the magnitude and color conditions to be classified as this type in the catalog of red sources from \citet{Robitaille_2008}. The title includes its VIRAC2 source ID, period, and amplitude.}
        \label{example_contra}%
\end{figure}

Considering the AGBs/LPVs and YSOs discussed above, the agreement fraction is approximately $63\%$ between the catalog of red sources and this study. However, examining the light-curve amplitudes of sources classified as YSOs in the catalog of red sources but identified as AGBs in our catalog reveals variability signatures characteristic of the AGB evolutionary stage. An illustrative example is shown in Fig.~\ref{example_contra}, where the source exhibits a periodic and nearly sinusoidal modulation. Its period of $\sim135\,\rm days$ and $K_\mathrm{s}$ band amplitude of $0.85\,\rm mag$ are inconsistent with variability driven by rotational modulation in a young star. This further highlights that photometric magnitudes and colors alone cannot robustly disentangle these populations.

\subsection{Validation of our classification}
\label{sec:validation}

To better analyze the discrepancy mentioned above, we used a simple probabilistic approach. Considering our final classifications as a reference, we took the colors $[8.0]-[24]$, $[4.5]-[24]$, $J-K_{\rm s}$, $H-K_{\rm s}$, and $J-H$, along with the $[4.5]$ magnitude, robust $\Delta K_{\rm{s}}$, $\log P$, $Q$, and $M$ (photometry+variability model) values to train a logistic-regression classifier \citep{Cox_1958, Pedregosa_2011} using stratified five-fold cross-validation, so that each source is assigned an out-of-fold probability of being an LPV, $P({\rm LPV})$, estimated only by models that did not include that source during training. The \citet{Robitaille_2008} criteria were used only to define the photometric agreement and disagreement groups, not as classifier labels. All logistic-regression models in this section and in Appendix~\ref{app:validation}, where we further describe different procedures, share the same implementation. In all of them, features were standardized (zero mean, unit variance) using a scaler fit independently within each cross-validation training fold (via a scikit-learn \texttt{Pipeline}). We used the \texttt{liblinear} solver with the default L2 regularization strength ($C=1$); \texttt{class\_weight="balanced"} was applied in every model to correct for the unequal class sizes; the classification threshold was $P(\mathrm{LPV})=0.5$; and a fixed random seed (\texttt{random\_state=42}) was used for the stratified $5$-fold splits.\\

This full (photometry+variability) model achieves an out-of-fold balanced accuracy of $97.9\%$ (ordinary accuracy $99.3\%$, $ \rm{ROC-AUC}=0.9997$) over the $N=150$ common sources with \citet{Robitaille_2008} that have photometric measurements in all Spitzer bands and in $J$, $H$ and $K_{\rm s}$, since the model also uses the colours $J-K_{\rm s}$, $H-K_{\rm s}$ and $J-H$. This almost perfect separation is expected, since $\log P$ and robust $\Delta K_{\rm s}$ are themselves close to the variables defining our LPV/YSO selection, and should be read as a sanity check. For a stricter internal check, we considered features independent of the period-amplitude selection. We defined a model using only the IR photometric colors (i.e., excluding the variability indicators or photometry-only model). This achieves a balanced accuracy of $63.6\%$ ($\rm{ROC-AUC}=0.683$; ordinary accuracy $58.7\%$) on the same $N=150$ sources (Table~\ref{tab:confusion_photometry}). We report the balanced accuracy and ROC-AUC here because the sample is imbalanced ($24$ YSOs versus $126$ LPVs). This confirms that near- and mid-IR colors alone carry modest, but genuine, separating power, consistent with the overlap seen in Fig.~\ref{cmd_spitzer} and Fig.~\ref{robitaille}.\\

\begin{table}
\small
\centering
\caption{Out-of-fold confusion matrix for the photometry-only
logistic-regression classifier ($[4.5]$, $[8.0]-[24]$, $[4.5]-[24]$, $J-K_{\rm s}$, $H-K_{\rm s}$, $J-H$)}
\label{tab:confusion_photometry}
\begin{tabular}{lccc}
\hline\hline
True class & Pred. YSO & Pred. LPV & $N$ \\
\hline
YSO      & 17 & 7  & 24  \\
LPV  & 55 & 71 & 126 \\
\hline
\end{tabular}
\tablefoot{$N=150$ common sources with \citet{Robitaille_2008},
stratified 5-fold cross-validation. Accuracy $58.7\%$, $\rm{ROC-AUC} =0.683$.}
\end{table}

For the $94$ sources that we classified as LPVs but that are in Robitaille's YSO region, all of them among the $126$ LPVs of the $N=150$ sample, we found a median probability of $P({\rm LPV})=0.94$ using the full (photometry+variability) model, indicating they are consistent with our classification, as expected. The photometry-only model alone remains ambiguous for these sources (median $P(\rm LPV)=0.39$--$0.41$). In contrast, the $9$ sources classified as YSOs in our study that lie in Robitaille's AGB region have a median $P({\rm LPV})=9.4\times10^{-4}$ with the full model, consistent with our classification. Again, because period and amplitude are themselves close to the variables used to define our LPV/YSO selection, the near-perfect separation of the full model should not be read as fully independent validation. The photometry-only result above, which uses no information from our own selection criteria, is the stricter test.\\

To perform additional validation tests that could be more independent of our own classification procedure, we first isolated the variability parameters $Q$ and $M$ from period and amplitude. As a second test, we also quantified the classifier performance as a function of $K_{\rm s}$ magnitude, robust $\Delta K_{\rm s}$, and period. We estimated the explicit precision and recall for both stellar evolutionary stages. All these are described in Appendix~\ref{app:validation}. In summary, a classifier trained only on $Q$ and $M$ (free of the burster/dipper selected sources) achieves a balanced accuracy of $59.1\%$ ($\rm{ROC-AUC}=0.63$; ordinary accuracy = $58\%$). Among the $44$ YSOs in the period-amplitude region shared with semi-regular variables in the literature, $32/44$ ($73\%$) are correctly classified. Classification accuracy degrades toward faint magnitudes; across amplitude and period tertiles it follows the class composition of each tertile rather than isolating the overlap region. We stress that all classifiers described here and in Appendix~\ref{app:validation} are used solely as internal-consistency checks, and none of them alter any of the final classifications reported in the present work. A genuinely external validation, using sources that never entered our training or visual inspection, is presented in Appendix~\ref{app:validation_external}: the classifier recovers $69.8\%$ [$95\%$ CI: $60.0$--$78.1\%$] of the $96$ emission-line YSOs of \citet{Borissova_2025} with a measured period, and assigns none of the external Miras to the YSO class, so that the precision of the predicted-YSO class is $100\%$ ($67/67$, $95\%$ CI $94.6$--$100\%$). Further, we also note that only spectroscopic surveys and follow-ups of these possible contaminants, such as the forthcoming MOONS and 4MOST, will be the main tools to address this correctly.\\


\subsection{LPV contamination in Galactic star formation estimates}

To place this possible contamination level in the context of Galactic star formation rate estimates, we computed the contamination fraction, $f_{\rm cont}(l) = N_{\rm LPV}(l)/N_{\rm matched}(l)$, in the same $10^\circ$-wide longitude bins as Fig.~\ref{spatial}, where $N_{\rm matched}(l)$ is the full VIRAC2$\times$SPICY cross-matched sample ($58\,737$ sources). The contamination fraction is $f_{\rm cont}=1.26\%$ ($95\%$ Wilson CI: $1.18$--$1.36\%$). This contamination fraction is significantly higher toward the Bulge ($|l|\leq10^\circ$: $373$ LPVs among $22\,424$ matched sources, $f_{\rm cont}=1.66\%$, 95\% CI $1.50$--$1.84\%$) than in the disk ($|l|>10^\circ$: $369$ LPVs among $36\,313$ matched sources, $f_{\rm cont}=1.02\%$, 95\% CI $0.92$--$1.12\%$). A two-sided Fisher's exact test comparing these two Galactic regions confirms that this difference is modest but statistically robust (odds ratio$=1.65$, $p=1.8\times10^{-11}$), and confirms quantitatively the trend already visible in the spatial distributions of Fig.~\ref{spatial}.\\

Under the simple approximation that raw star formation rate estimates scale linearly with YSO counts, this contamination fraction translates directly into $(\mathrm{SFR}_{\rm raw}-\mathrm{SFR}_{\rm corrected})/\mathrm{SFR}_{\rm raw} = f_{\rm cont}(l)$, independent of the initial mass function, mean stellar mass, and YSO lifetime, which are typically required to convert counts into a star formation rate and are beyond the scope of the present work. Thus, counting all VIRAC2$\times$SPICY sources as true YSOs would overestimate the inferred star formation rate by exactly the fractions quoted above: $1.26\%$ on average, $1.66\%$ toward the Bulge and $1.02\%$ in the disk. Nevertheless, our LPV and YSO catalogs are not intended to be comprehensive. Saturated bright sources, LPVs with periods beyond the VIRAC2 baseline, and irregular or aperiodic light curves excluded by our selection criteria can affect their completeness. As with the LPV and YSO catalogs themselves, $f_{\rm cont}$ should therefore be considered a lower limit on the true contamination of the SPICY catalog.


\section{Conclusion}

Starting from a sample of $7616$ LPV candidates, we proposed $742$ sources as high-confidence LPV contaminants among the SPICY catalog. We excluded most LPV candidates from this list because of missing data. However, this does not mean they are not LPVs, only that they are not high-confidence LPVs. Despite representing a small fraction of the total entries in the catalog ($\sim1.3\%$), the consequences of removing them range from saving significant telescope time in follow-up observations to reducing biases in Galactic star formation rate estimates, particularly towards the Bulge. This number is almost $2.9$ times larger than the one obtained by \citet{Jheonn_2026} for the potential AGB stars in the SPICY catalog. At the same time, we identified $307$ candidates whose period, amplitude, and light curve shape provide additional evidence for their YSO nature.\\

Since light curve analysis has proven effective at separating these two stellar populations, the primary source of potential misclassifications comes from sources in the region where YSOs overlap with semi-regular variables among the LPVs. Nevertheless, for the majority of these sources, their spatial distribution, IR colors $H-K_{\rm{s}}$, $[3.6]-[4.5]$, $[5.8]-[8.0]$, $J-H$, lack of correlation between $W2$ and robust $\Delta K_{\rm{s}}$ amplitudes, and positions relative to the CTTS locus are more consistent with a YSO interpretation than with an LPV origin. Still, their YSO status should be treated with caution until confirmed by other means (spectroscopically, for instance).\\

As a ground-based survey, VVVX/VIRAC2 is limited by atmospheric effects that restrict our ability to identify every misclassified source \citep{Smith_2025}. Spectroscopic follow-up is required to confirm our labels; to our knowledge, the literature provides no spectra for most of these stars. Four of them are in the \textit{Gaia}-ESO Survey \citep{Gilmore_2022}, all showing lithium absorption ($\lambda \approx 670.8\,\rm nm$). Three of our LPVs are in the Apache Point Observatory Galactic Evolution Experiment \citep[APOGEE,][]{SDSS_2026} catalog, but none of them had reliable atmospheric parameters. Future facilities such as Vera Rubin \citep{Ivezic_2019} will extend variability analysis to less embedded objects, though heavily extincted stars will remain inaccessible without IR monitoring. With no equivalent IR survey planned, VVVX \citep{minniti+2010} remains among the most powerful tools for separating YSOs from evolved red variables. Identifying LPV contaminants in spatially unbiased catalogs such as SPICY is therefore essential for robustly characterizing star formation across the Galaxy.






\begin{acknowledgements}
      We thank the anonymous referee for their suggestions, which improved our paper. This work was supported by the Comité Mixto ESO -- Chile Postdoctoral Fellowship, the ANID BASAL Center for Astrophysics and Associated Technologies (CATA) through grant FB210003, and by FONDECYT Regular grant No. 1230731. COH acknowledges Agencia Nacional de Investigación y Desarrollo (ANID) through FONDECYT postdoctoral grant 3260854. AB acknowledges support under Germany’s Excellence Strategy through the Cluster of Excellence ORIGINS EXC–2094–390783311. VGV acknowledges support from ANID -- Millennium Science Initiative Program -- Center Code NCN2024\_001 and from FONDECYT Regular 1221352. 
\end{acknowledgements}

\bibliographystyle{aa} 
\bibliography{Wd1_bibliography.bib} 

\begin{appendix}
\nolinenumbers





\FloatBarrier 
\twocolumn

\section{Source selection summary}

Tables \ref{tab:period_search_LPV} and \ref{tab:period_search_YSO} show a summary for the number of sources that are included in each step, and their requirements or cuts, of the entire selection process for LPV and YSO candidates, respectively.

\begin{table*}[h!]
\caption{Summary of the main source-selection and period-search for the LPV sample.}
\label{tab:period_search_LPV}
\centering
\begin{tabular}{lll}
\hline\hline
Step & Selection / requirement & N \\
\hline
VIRAC2 x SPICY & Best match within $1\,\rm arcsec$ in the VIRAC2 source catalog & 58\,737 \\
Light curve quality & \texttt{ast\_res\_chisq}$<30$, \texttt{chi}$<5$ & 54\,580 \\
Period-search input & $N(K_\mathrm{s})\geq50$ epochs; $\sigma(K_\mathrm{s})/\Delta K_\mathrm{s,raw}\leq0.3$, two non-alias peaks & 45\,499 \\
Finite $Q(P_1)$ and $Q(P_2)$ & Finite and physical periodicity ($0<Q<1$) for $P_i$ & 37\,046 \\
Periodic and symmetric & $0\leq Q\leq0.6$, $-0.4\leq M\leq0.4$ & 15\,010 \\
Symmetric LPV candidates & $\Delta K_\mathrm{s}\geq0.35\,\rm mag$, $P\geq80$ d & 7616 \\
Pre-filtered symmetric LPV candidates & period boundary, $N_\mathrm{epochs}\geq100$, robust $\Delta K_\mathrm{s}\geq0.35\,\rm mag$ & $2785$ \\
Final symmetric LPV candidates  & visually confirmed & $559$ \\
Burster/dipper LPV candidates\tablefootmark{a} & $|M|>0.4$, $0\leq Q\leq0.6$, $P\geq80\,\rm days$ & $7204$ \\
Pre-filtered burster/dipper LPV candidates & $|M|>0.4$, same $Q$/$P$ box and screening as LPV & $2800$ \\
Final Burster/dipper LPVs & visually confirmed and $Q\leq0.3$ & $183$ \\
Final LPV catalog & final symmetric and final burster/dipper candidates & $742$ \\
\hline
\end{tabular}
\tablefoot{
\tablefoottext{a}{This branch does not descend from the periodic-and-symmetric row, which requires $|M|\leq0.4$; it separates earlier, from the sources with finite and physical $Q(P_1)$ and $Q(P_2)$. No amplitude cut is applied at this stage: the $\Delta K_{\rm s}\geq0.35$ mag requirement enters only in the subsequent pre-filtering, and as the robust amplitude.}
}
\end{table*}

\begin{table*}[h!]
\caption{Summary of the main source-selection and period-search for the YSO sample.}
\label{tab:period_search_YSO}
\centering
\setlength{\tabcolsep}{3.2pt}
\begin{tabular}{lll}
\hline\hline
Step & Selection / requirement & N \\
\hline
Periodic and symmetric & $0\leq Q\leq0.6$, $-0.4\leq M\leq0.4$ & 15\,010 \\
Symmetric YSO candidates & $\Delta K_\mathrm{s}<0.4\,\rm mag$, $P\leq50\,\rm days$ & $2004$ \\
Final YSO catalog & visually confirmed & $307$ \\
\hline
\end{tabular}
\end{table*}

\begin{table*}
\centering
\caption{Columns list the VIRAC2 source identifier, equatorial coordinates RA and Dec, $JHK_{\rm{s}}$ magnitudes, period $P$ in
days, raw $K_{\rm s}$ amplitude, periodicity $Q$, asymmetry $M$, and the false alarm probability of the selected period, FAP.}
\label{tab:lpv_sample}
\begin{tabular}{ccccccccccc}
\hline\hline
Source ID & RA & Dec & $J$ & $H$ & $K_{\rm s}$ & $P$ & $\Delta K_s$ & $Q$ & $M$ & FAP \\
 & [deg] & [deg] & [mag] & [mag] & [mag] & [d] & [mag] & & & \\
\hline

13468636002418 & 266.88 & $-$28.30 & \ldots & \ldots & 14.60 & 480.88 & 1.20 & 0.04 & $-$0.28 & $2.37 \times 10^{-150}$ \\
13468635010894 & 266.80 & $-$28.27 & \ldots & 16.91 & 12.79 & 440.66 & 1.17 & 0.06 & 0.11 & $1.48 \times 10^{-147}$ \\
13538257015414 & 265.88 & $-$29.01 & \ldots & 15.76 & 12.63 & 507.45 & 2.24 & 0.07 & $-$0.29 & $9.07 \times 10^{-165}$ \\
13399011000357 & 267.45 & $-$27.58 & \ldots & \ldots & 14.58 & 840.64 & 0.47 & 0.07 & $-$0.01 & $5.18 \times 10^{-55}$ \\
13636559008835 & 265.71 & $-$30.02 & \ldots & 15.25 & 12.07 & 456.51 & 1.22 & 0.07 & $-$0.27 & $8.18 \times 10^{-224}$ \\
\hline
\end{tabular}
\end{table*}


\section{List of periodic YSOs}
VVVX - VIRAC2 photometry and periodic properties of the YSO sample. The full catalog is provided as machine-readable supplementary material.
\begin{table*}[h]
\caption{Columns list the VIRAC2 source identifier, equatorial coordinates RA and Dec, $JHK_\mathrm{s}$ magnitudes, rotation period $P$ in days, raw $K_\mathrm{s}$ amplitude, periodicity $Q$, asymmetry $M$ and the false alarm probability of the selected period $P$, FAP.}
\label{YSOs_table}
\centering
\begin{tabular}{lcccccccccc}
\hline\hline
Source ID &
RA &
Dec &
$J$ &
$H$ &
$K_\mathrm{s}$ &
$P$ &
$\Delta K_\mathrm{s}$ &
$Q$ &
$M$ &
FAP \\
 &
[deg] &
[deg] &
[mag] &
[mag] &
[mag] &
[d] &
[mag] &
 &
 &
 \\
\hline
15003220006138 & 249.98 & $-$45.86 & \ldots & 15.51 & 13.57 & 18.57 & 0.12 & 0.02 & 0.05 & $5.10 \times 10^{-6}$ \\
15319198000034 & 244.80 & $-$50.20 & 17.92 & 14.64 & 13.09 & 19.45 & 0.16 & 0.02 & 0.11 & $2.30 \times 10^{-9}$ \\
15092483005085 & 250.15 & $-$47.06 & \ldots & \ldots & 13.33 & 0.86 & 0.18 & 0.02 & $-$0.21 & $1.31 \times 10^{-2}$ \\
16030791001768 & 212.03 & $-$61.77 & \ldots & 16.77 & 14.76 & 3.25 & 0.26 & 0.03 & $-$0.24 & $4.79 \times 10^{-4}$ \\
15871723003438 & 228.93 & $-$58.88 & 13.41 & 12.39 & 11.99 & 1.28 & 0.13 & 0.03 & 0.18 & $5.66 \times 10^{-5}$ \\
\hline
\end{tabular}
\end{table*}

\section{Additional classifier validation tests}
\label{app:validation}

To independently assess the classification of sources (YSO versus LPV) beyond the period-amplitude selection, we only considered the light curve metrics $Q$ and $M$. We evaluated them for the $44$ YSOs located within the period-amplitude region shared with semi-regular variables in the literature ($20 \le P \le 50\,\rm days$, $\Delta K_{\rm s} \lesssim 0.4\,\rm mag$), as well as for the remainder of the sample. Of the $742$ LPVs, $183$ were classified as burster/dipper based on their $M$ values and were exclusively selected through visual inspection with $Q\le0.3$. To avoid selection bias, we trained the model using only the $559$ symmetric LPVs (those without the $Q\leq 0.3$ criterion) and all $307$ YSOs. A logistic-regression model utilizing only $Q$ and $M$ achieved a balanced accuracy of $59.1\%$ ($\rm{ROC-AUC}=0.63$; ordinary accuracy $58\%$) across these $866$ sources ($559$ symmetric LPVs and $307$ YSOs). In comparison, a majority-class baseline would yield $64.5\%$ ordinary accuracy. These results indicate that periodicity ($Q$) and asymmetry ($M$) alone provide limited discriminative power for this sample, with $Q$ being the dominant factor (standardized coefficient $-0.40$, compared to $-0.18$ for $M$). When focusing exclusively on the $44$ YSOs in the semi-regular overlap region, the $Q$-and-$M$-only model correctly classifies $32$ of $44$ ($73\%$) sources, comparable to its overall performance. This finding aligns with the observation that the median $Q$ of symmetric LPVs ($\sim0.39$) remains somewhat lower than that of YSOs ($\sim0.49$), suggesting a modest tendency toward more coherent, periodic light curves in LPVs relative to the spot-modulated and potentially accretion-affected variability observed in YSOs.\\

We also investigated the classifier performance as a function of $K_{\rm s}$ magnitude, robust $\Delta K_{\rm s}$, and period, using the out-of-fold predictions of the $Q$-and-$M$-only model, which is independent of these three quantities. We evaluated its performance in magnitude, amplitude, and period tertiles. We found that its accuracy declines modestly from bright to faint sources ($61\%\to58\%\to55\%$), consistent with increased photometric uncertainty degrading $Q$ sensitivity for faint stars. For the robust amplitude and period, accuracy is highest at the extremes ($63$--$68\%$) and lowest in the intermediate bin ($\Delta K_{\rm s}\sim0.27$--$0.55$ mag, $P\sim40$--$260$ days; $42$--$44\%$). This behavior follows from the sample definition, since amplitude and period are the quantities that separate our two classes, so their tertiles nearly coincide with the class label. The lowest and highest tertiles contain a single class each, $289$ YSOs and $289$ LPVs, so their accuracy is the recall of that class, while the intermediate tertile holds $270$ LPVs against $18$ YSOs. Only one source falls in both the intermediate tertile and the overlap region defined above, so this breakdown describes how performance varies across the selection variables and does not isolate the ambiguous region.\\

We summarize the out-of-fold precision and recall for both classes in Table~\ref{tab:purity_completeness}, for all the mentioned tests that considered auxiliary classifiers built on information independent of our selection procedure or period-amplitude information, together with the \citet{Robitaille_2008} color magnitude criterion evaluated against our final classification as an external reference. We emphasize that these quantities should not be interpreted as direct measurements of the astrophysical purity or completeness of the final LPV/YSO catalogs, since the training labels used to compute them are our own adopted visual classifications (Sect.~\ref{sec:visual_inspection}). These tests should therefore be read as internal consistency checks, rather than as external validation. The genuinely external validation, using independently confirmed sources from catalogs never involved in our selection or training, is presented separately in Appendix~\ref{app:validation_external}.\\

\begin{table*}
\centering
\caption{Out-of-fold precision and recall of auxiliary classifiers relative to our adopted visual classifications, using information independent of our period-amplitude selection.}
\label{tab:purity_completeness}
\begin{tabular}{lcccc}
\hline\hline
Test & Prec.(LPV) & Recall(LPV) & Prec.(YSO) & Recall(YSO) \\
\hline
Photometry-only ($N=150$)     & 0.91 & 0.56 & 0.24 & 0.71 \\
$Q$+$M$-only ($N=866$)        & 0.73 & 0.55 & 0.44 & 0.63 \\
Robitaille's own color cut\tablefootmark{a} & 0.94 & 0.62 & 0.18 & 0.70 \\
\hline
\end{tabular}
\tablefoot{
\tablefoottext{a}{\citet{Robitaille_2008} color-magnitude
criterion evaluated against our final classification as reference
($N=277$); shown for comparison, not as our own classifier.}
}
\end{table*}

From Table~\ref{tab:purity_completeness}, we note that the precision for the YSO class is consistently lower ($0.18$--$0.44$) than for the LPV class ($0.73$--$0.94$) across all three tests. Although each classifier used class-balanced weighting to avoid biasing the decision boundary toward the majority class, and the estimated recall of each class is comparable between the two populations (e.g., $55\%$ for LPVs versus $63\%$ for YSOs in the $Q$-and-$M$-only test), the precision for the minority YSO class remains lower simply because LPVs substantially outnumber YSOs in all three samples (by factors of $\sim 1.8$ to $\sim 8$, depending on the test). Even a moderate per-source misclassification rate among the much larger LPV candidate sample contributes a comparable or larger absolute number of false positives to the predicted-YSO sample than the number of true YSOs. This reflects the limited discriminating power of photometry-only or $Q$-and-$M$-only tests for the specific subset of sources where the two classes overlap, combined with the unequal sizes of the two populations, consistent with the qualitative overlap already discussed in Fig.~\ref{cmd_spitzer} and Fig.~\ref{robitaille}.\\

In addition, because our final candidate catalogs result from a combination of automated period-amplitude cuts and visual inspection rather than a single statistical threshold, we do not report a Poisson uncertainty on $N_{\rm LPV}=742$ or $N_{\rm YSO}=307$. We consider the precision and recall estimates above, together with the FAP statistics already reported for the adopted periods in Tables \ref{tab:lpv_sample} and \ref{YSOs_table}, to be the most meaningful measure of the reliability of our final catalogs given the nature of the selection process.

\section{Validation with other LPV and YSO catalogs}
\label{app:validation_external}

As a fully independent external test, we applied period-only and period$+$amplitude logistic-regression models (trained on the complete local sample of $742$ LPVs and $307$ YSOs) to external, spectroscopically or independently confirmed samples. We included emission-line member YSOs from \citet{Borissova_2025} ($N=112$), and Mira/LPV candidates from the OGLE \citep[$N=65\,981$;][]{Iwanek_2022}, Albarrac\'in et al.\ \citep[$N=3602$;][]{Albarracin_2025}, and Nikzat et al.\ \citep[$N=130$;][]{Nikzat_2022} catalogs. We discarded common sources or matches within $1''$ of any of the local training sources (there being $0$, $4$, $205$, and $12$ such sources for the four catalogs, respectively, the last two numbers representing the overlap between the VVV-based Mira searches and our own survey area), so that no external source used for validation appeared in the training sample.\\

Using period alone, our classifier recovers $69.8\%$ [95\% Wilson CI: $60.0$--$78.1\%$] of the $96$ external Borissova YSOs with a measured period (out of $112$ strict emission-line members), while recovering $100\%$ of the external Miras/LPVs in all three catalogs, so that all $67$ sources assigned to the YSO class are genuine YSOs (precision $100\%$, $95\%$ Wilson CI $94.6$--$100\%$). The latter is expected given that confirmed Miras have periods far into the long-period regime, and should be read as a comparatively easy test. The period$+$amplitude test cannot be applied to all three catalogs and was therefore evaluated pairwise, the external YSO catalog by \citet{Borissova_2025} against one of the two external LPV catalogs with $K_{\rm s}$ amplitude measurements at a time. The OGLE catalog of \citet{Iwanek_2022} is an optical survey and reports no $K_{\rm s}$ amplitude, so none of its $65\,977$ sources can enter a test that uses $\Delta K_{\rm s}$ as a feature. Adding $K_{\rm s}$-band amplitude changes YSO completeness only marginally ($66.7\%$ [$56.8$--$75.3\%$]). LPV completeness remains $100\%$ for the \citet{Nikzat_2022} sample, with no false positives. Still, it drops to $92.2\%$ ($3133/3397$, $95\%$ CI $91.3$--$93.1\%$) for \citet{Albarracin_2025}, plausibly reflecting the different $K_{\rm s}$ amplitude definition used in that catalog compared to our robust $\Delta K_{\rm s}$. In that pairing, $264$ of the $328$ sources assigned to the YSO class are external Miras, so the precision of the predicted YSO class falls to $19.5\%$ ($64/328$, $95\%$ Wilson CI $15.6$--$24.1\%$). As with the internal precision estimates in Table~\ref{tab:purity_completeness}, this mainly reflects the highly unequal external sample sizes in that particular pairing ($3397$ Miras versus $96$ YSOs), combined with a per-source misclassification rate of only $7.8\%$ on the \citet{Albarracin_2025} sample, not evidence of a fundamentally unreliable classifier: the same model evaluated against the $118$ \citet{Nikzat_2022} Miras, where the two samples are comparable in size, yields no false positives. Overall, this external validation, independent of our classification procedure, broadly matches the internal cross-validation results in Appendix~\ref{app:validation}, supporting the general reliability of our period-amplitude-based classification outside our sample.\\

Finally, we also investigated whether the $29$ Borissova YSOs misclassified as LPVs in the period-only test reflect unreliable periods rather than a genuine classifier limitation. \citet{Borissova_2025} define a source as RELIABLE if $K_{\rm s}\geq11.5$ and $N_{\rm obs}\geq30$, and further classify a source as periodic if it is RELIABLE and has $\rm{FAP}<0.2$. Under this periodicity criterion, $18$ of the $29$ sources remain periodic. Because periods close to the $500$-day upper boundary of their Lomb--Scargle search are more likely to be spurious grid solutions, we additionally imposed a more conservative cut of our own, requiring $\rm{FAP}<0.05$ and excluding periods within $10$ days of the search boundary. After these cuts, we considered only $10$ of the $29$ sources reliably periodic. This indicates that a meaningful fraction, but not all, of the nominal long periods in this subsample reflect unreliable solutions rather than genuine periodicity. For these remaining $10$ sources with robust long periods under our stricter criterion, we note that rotational periods of $\sim100$--$500\,\rm days$ are physically unlikely for pre-main-sequence stars, which typically show spot-modulated rotational periods of only a few to tens of days \citep{Rebull_2014}. Since the \citet{Borissova_2025} light curves are not publicly available, we cannot determine the physical origin of these apparently robust long periods and we speculate that they may instead trace longer timescale physical processes (e.g., periodic occultation by disk substructures, orbital motion in an unresolved binary) rather than stellar rotation. These objects illustrate an important limitation of period-based classification: genuine YSOs with long-timescale variability unrelated to stellar rotation can occupy the same period regime as LPVs and may therefore be misclassified by our method.

\end{appendix}
\end{document}